\documentclass[preprintnumbers,groupedaddress,nofootinbib,aps]{revtex4}

\usepackage{multirow}
\usepackage{amsmath,amssymb}
\usepackage{graphicx}
\usepackage{color}
\usepackage{hyperref}
\usepackage{url}
\usepackage{slashed}
\usepackage{subfigure}
\usepackage[usenames,dvipsnames]{xcolor}

\newcommand{\highscale}{k}
\newcommand{\mh}{m_h}

\begin{document}
\def\contentsname{{\normalsize Content}}
\def\tablename{Table}
\def\figurename{Figure}

\def\pveto{P_\text{veto}}
\def\nj{n_\text{jets}}
\def\meff{m_\text{eff}}
\def\ptmin{p_T^\text{min}}
\def\gtot{\Gamma_\text{tot}}
\def\as{\alpha_s}
\def\az{\alpha_0}
\def\gz{g_0}
\def\w{\vec{w}}
\def\sdag{\Sigma^{\dag}}
\def\s{\Sigma}
\newcommand{\psib}{\overline{\psi}}
\newcommand{\Psib}{\overline{\Psi}}
\newcommand\one{\leavevmode\hbox{\small1\normalsize\kern-.33em1}}
\newcommand{\Mpl}{M_\mathrm{Pl}}
\newcommand{\p}{\partial}
\newcommand{\mat}{\mathcal{M}}
\newcommand{\lag}{\mathcal{L}}
\newcommand{\ord}{\mathcal{O}}
\newcommand{\ope}{\mathcal{O}}
\newcommand{\qqquad}{\qquad \qquad}
\newcommand{\qqqquad}{\qquad \qquad \qquad}

\newcommand{\qb}{\bar{q}}
\newcommand{\matx}{|\mathcal{M}|^2}
\newcommand{\really}{\stackrel{!}{=}}
\newcommand{\msbar}{\overline{\text{MS}}}
\newcommand{\qns}{f_q^\text{NS}}
\newcommand{\lqcd}{\Lambda_\text{QCD}}
\newcommand{\met}{\slashchar{p}_T}
\newcommand{\pmiss}{\slashchar{\vec{p}}_T}

\newcommand{\sq}{\tilde{q}}
\newcommand{\go}{\tilde{g}}
\newcommand{\st}[1]{\tilde{t}_{#1}}
\newcommand{\stb}[1]{\tilde{t}_{#1}^*}
\newcommand{\nz}[1]{\tilde{\chi}_{#1}^0}
\newcommand{\cp}[1]{\tilde{\chi}_{#1}^+}
\newcommand{\cm}[1]{\tilde{\chi}_{#1}^-}
\newcommand{\CP}{CP}

\providecommand{\mg}{m_{\tilde{g}}}
\providecommand{\mst}{m_{\tilde{t}}}
\newcommand{\msn}[1]{m_{\tilde{\nu}_{#1}}}
\newcommand{\mch}[1]{m_{\tilde{\chi}^+_{#1}}}
\newcommand{\mne}[1]{m_{\tilde{\chi}^0_{#1}}}
\newcommand{\msb}[1]{m_{\tilde{b}_{#1}}}
\newcommand{\vsm}{\ensuremath{v_{\rm SM}}}

\newcommand{\mev}{{\ensuremath\rm MeV}}
\newcommand{\gev}{{\ensuremath\rm GeV}}
\newcommand{\tev}{{\ensuremath\rm TeV}}
\newcommand{\fb}{{\ensuremath\rm fb}}
\newcommand{\ab}{{\ensuremath\rm ab}}
\newcommand{\pb}{{\ensuremath\rm pb}}
\newcommand{\sign}{{\ensuremath\rm sign}}
\newcommand{\ifb}{{\ensuremath\rm fb^{-1}}}
\newcommand{\ipb}{{\ensuremath\rm pb^{-1}}}

\def\slashchar#1{\setbox0=\hbox{$#1$}           
   \dimen0=\wd0                                 
   \setbox1=\hbox{/} \dimen1=\wd1               
   \ifdim\dimen0>\dimen1                        
      \rlap{\hbox to \dimen0{\hfil/\hfil}}      
      #1                                        
   \else                                        
      \rlap{\hbox to \dimen1{\hfil$#1$\hfil}}   
      /                                         
   \fi}
\newcommand{\dslash}{\slashchar{\partial}}
\newcommand{\Dslash}{\slashchar{D}}

\newcommand{\eg}{\textsl{e.g.}\;}
\newcommand{\ie}{\textsl{i.e.}\;}
\newcommand{\etal}{\textsl{et al}\;}

\setlength{\floatsep}{0pt}
\setcounter{topnumber}{1}
\setcounter{bottomnumber}{1}
\setcounter{totalnumber}{1}
\renewcommand{\topfraction}{1.0}
\renewcommand{\bottomfraction}{1.0}
\renewcommand{\textfraction}{0.0}
\renewcommand{\thefootnote}{\fnsymbol{footnote}}

\newcommand{\rig}{\rightarrow}
\newcommand{\lrig}{\longrightarrow}
\renewcommand{\d}{{\mathrm{d}}}
\newcommand{\be}{\begin{eqnarray*}}
\newcommand{\ee}{\end{eqnarray*}}
\newcommand{\gl}[1]{(\ref{#1})}
\newcommand{\ta}[2]{ \frac{ {\mathrm{d}} #1 } {{\mathrm{d}} #2}}
\newcommand{\bee}{\begin{eqnarray}}
\newcommand{\eee}{\end{eqnarray}}
\newcommand{\beeq}{\begin{equation}}
\newcommand{\eeeq}{\end{equation}}
\newcommand{\mc}{\mathcal}
\newcommand{\mr}{\mathrm}
\newcommand{\ep}{\varepsilon}
\newcommand{\emt}{$\times 10^{-3}$}
\newcommand{\emfo}{$\times 10^{-4}$}
\newcommand{\emfi}{$\times 10^{-5}$}

\newcommand{\revision}[1]{{\bf{}#1}}

\newcommand{\hzero}{h^0}
\newcommand{\Hzero}{H^0}
\newcommand{\Azero}{A^0}
\newcommand{\PHiggs}{H}
\newcommand{\PW}{W}
\newcommand{\PZ}{Z}

\newcommand{\sw}{\ensuremath{s_w}}
\newcommand{\cw}{\ensuremath{c_w}}
\newcommand{\swd}{\ensuremath{s^2_w}}
\newcommand{\cwd}{\ensuremath{c^2_w}}

\newcommand{\mhhd}{\ensuremath{m^2_{\Hzero}}}
\newcommand{\mhh}{\ensuremath{m_{\Hzero}}}
\newcommand{\mlhd}{\ensuremath{m^2_{\hzero}}}
\newcommand{\Mlh}{\ensuremath{m_{\hzero}}}
\newcommand{\mad}{\ensuremath{m^2_{\Azero}}}
\newcommand{\mhpd}{\ensuremath{m^2_{\PHiggs^{\pm}}}}
\newcommand{\mhp}{\ensuremath{m_{\PHiggs^{\pm}}}}

 \newcommand{\sa}{\ensuremath{\sin\alpha}}
 \newcommand{\ca}{\ensuremath{\cos\alpha}}
 \newcommand{\cad}{\ensuremath{\cos^2\alpha}}
 \newcommand{\sad}{\ensuremath{\sin^2\alpha}}
 \newcommand{\sbd}{\ensuremath{\sin^2\beta}}
 \newcommand{\cbd}{\ensuremath{\cos^2\beta}}
 \newcommand{\cb}{\ensuremath{\cos\beta}}
 \renewcommand{\sb}{\ensuremath{\sin\beta}}
 \newcommand{\tanbd}{\ensuremath{\tan^2\beta}}
 \newcommand{\cotbd}{\ensuremath{\cot^2\beta}}
 \newcommand{\tanb}{\ensuremath{\tan\beta}}
 \newcommand{\tb}{\ensuremath{\tan\beta}}
 \newcommand{\cotb}{\ensuremath{\cot\beta}}

\newcommand{\GeV}{\ensuremath{\rm GeV}}
\newcommand{\MeV}{\ensuremath{\rm MeV}}
\newcommand{\TeV}{\ensuremath{\rm TeV}}

\title{The Higgs Mass and the Scale of New Physics}

\author{Astrid Eichhorn$^{1,2}$,
        Holger Gies$^3$,
        Joerg Jaeckel$^4$,
        Tilman Plehn$^4$, 
        Michael M. Scherer$^4$,
        Ren\'e Sondenheimer$^3$}

\affiliation{$^1$ Perimeter Institute for Theoretical Physics, Waterloo, Ontario, Canada}
\affiliation{$^2$ Blackett Laboratory, Imperial College London, U.K.}
\affiliation{$^3$ Theoretisch-Physikalisches Institut, Universit\"at Jena, Germany}
\affiliation{$^4$ Institut f\"ur Theoretische Physik, Universit\"at Heidelberg, Germany}

\begin{abstract}
In view of the measured Higgs mass of 125 GeV, the perturbative
renormalization group evolution of the Standard Model suggests that
our Higgs vacuum might not be stable.  We connect the usual
perturbative approach and the functional renormalization group which
allows for a straightforward inclusion of higher-dimensional operators
in the presence of an ultraviolet cutoff.  In the latter framework we
study vacuum stability in the presence of higher-dimensional
operators.  We find that their presence can have a sizable influence
on the maximum ultraviolet scale of the Standard Model and the
existence of instabilities. Finally, we discuss how such operators can
be generated in specific models and study the relation between the
instability scale of the potential and the scale of new physics
required to avoid instabilities.

\end{abstract}

\maketitle

\bigskip \bigskip \bigskip
\tableofcontents
\newpage

\section{Introduction}
\label{sec:intro}

The recent discovery of a light, to current experimental precision
elementary Higgs boson~\cite{higgs,discovery} has a major effect on
the entire field of particle physics. A narrow Higgs boson with a mass
of $125~\gev$ allows for an extrapolation of all fundamental
interactions, except gravity, toward higher energy scales in terms of
perturbative gauge theories. In particular, the renormalization group
(RG) provides a link between the electroweak scale and more
fundamental, higher energy scales. Renormalization group
considerations for the Higgs sector typically provide two types of
limits. An upper limit on the Higgs mass arises from the Higgs
self-coupling becoming large (triviality bound)~\cite{triviality},
while a lower limit arises from stability considerations of the Higgs
potential~\cite{stability_orig,lhc_review}. The observed low Higgs
mass suggests that the triviality bound will not be a problem for
energy scales below the Planck scale. When it comes to the stability
of the Higgs potential the situation is less
clear~\cite{metastable_before,
  metastable_after,shaposhnikov2012,giudice,sm_only}.\bigskip

As for any potential we
refer to the Higgs potential as stable if it is
bounded from below.  We can attempt to link this condition to the
sign of the quartic Higgs self-coupling evaluated at different energy
scales~\cite{stability_orig}, but the details of such a connection
need to be tested carefully. Throughout this paper we will refer to
this link as the perturbative or perturbatively renormalizable
approach, because it typically relies on the assumption that all
operators in the Higgs Lagrangian have at most mass dimension
four. The great advantage of this approach is that the running
couplings are known to high precision in the perturbative Standard
Model, so we can compare its precise predictions to the measurements
in the Higgs sector.

Alternatively, we can ask the question which additional operators, for
example higher powers in the Higgs field, can render the Higgs vacuum
stable and how such effective operators can be linked to particle
physics models at the corresponding energy scales.  Whereas the
canonical irrelevance of higher-dimensional operators for low-energy
observables is natural from a Wilsonian viewpoint, the presence of
such operators is expected at high energies, once we approach the
scales of an underlying theory. Perturbatively, such operators can for
example appear as threshold effects from integrating out massive
fields. Alternatively, they can arise from other non-trivial effects
of a UV-completion.  As such higher-dimensional operators do not
appear in the usual, perturbatively renormalizable Higgs Lagrangian,
we refer to this generalized approach as non-perturbative. Its main
advantage is that it makes fewer assumptions about the Lagrangian at
energies significantly above our experimental reach.\bigskip

Based on the measured Higgs mass and top quark mass the perturbative
analysis suggests that the electroweak Higgs vacuum might indeed not
be stable. In the perturbative analysis the instability of the Higgs
vacuum can be linked to the energy scale at which the Higgs quartic
coupling turns negative. At the three-loop level this scale comes out
around $10^{10}~\gev$, where the precise value depends sensitively on
the Standard Model parameters.  Several physical interpretations of
this scale are possible: first, we can require that the Higgs
potential be stable at all scales. New degrees of freedom would then
have to appear below or at $10^{10}~\gev$, changing the
renormalization group evolution, and rendering the potential
stable~\cite{metastable_before,metastable_after}. We review one such
approach based on a Higgs portal with a scalar dark matter candidate
in the Appendix~\cite{eichhorn_scherer,bauer}.  Alternatively, we can
require that the absolute value of the (negative) Higgs quartic
coupling remains small. New physics only has to set in at the Planck
scale, tunneling rates to the true vacuum have to be small, and our
electroweak vacuum is meta-stable~\cite{giudice}.  The third, less
studied
option~\cite{gies2013,stable_frg,Gies:2014xha,Branchina:2005tu}
includes higher-dimensional operators which stabilize the Higgs vacuum
beyond $10^{10}~\gev$. Such operators will appear in theories with
physical cutoff or matching scales. We explore this third option. 

The higher-dimensional operators can be linked to particle physics
models, including heavy new particles \cite{werner}. Again, we choose
additional scalars coupled through a Higgs portal.  Unlike TeV-scale
new scalars, heavy particles do not rely on their renormalization
group running over a wide range of scales to affect the vacuum
structure.  Instead, they change the full set of running Higgs
self-couplings at high scales to modify physics at intermediate
scales.\bigskip

Our starting point is a field theory with higher-dimensional operators
but including only Standard Model fields.  This theory is
perturbatively non-renormalizable. Accordingly, there exists a finite
UV-cutoff $\Lambda$ in the general spirit of the Standard Model as an
effective theory. An appropriate tool for including all quantum
fluctuations in the presence of higher-dimensional operators as well as a
finite ultraviolet (UV)-cutoff is the functional renormalization
group~\cite{christof_eq}. Our discussion of the stability of the Higgs
potential proceeds in three steps:
\begin{enumerate}
\item To quantitatively study the stabilizing effects of
  higher-dimensional operators we need a model which reflects all
  essential features of the Standard Model. We construct and test such
  a model in Secs.~\ref{sec:sm} and~\ref{sec:model}.
\item In this toy model we describe the stability conditions, compute
  the possible Higgs mass range, and analyze its fixed-point structure
  (Secs.~\ref{sec:stability}-\ref{sec:fixed_point}).
\item Finally, in Sec.~\ref{sec:light} we study explicit models with
  additional heavy scalars and fermions and determine under which
  conditions they stabilize the Higgs vacuum, while yielding Higgs masses
  below the conventional stability bound.
\end{enumerate}
In Appendix~\ref{app:dark_matter} we discuss the effect of light
states, in Appendix~\ref{app:tunnel} we
summarize the computation of tunnelling rates, and in
Appendices~\ref{app:compare} and~\ref{app:mapping} we review the
impact of higher-dimensional operators and give a detailed link
between the perturbative and non-perturbative approach to the
Higgs--top renormalization group.

\section{Gauged Higgs--top model}\label{sec:gaugemodel}

The aim of this paper is to investigate 
Higgs mass bounds
and vacuum stability in the presence of higher-dimensional operators
and a finite ultraviolet cutoff.  In addition to the standard
perturbative running, we use functional renormalization group (FRG)
methods as a tool to compute the running of an extended set of
operators. Let us first set up a toy model that allows us to study the
essential features of the Standard Model in the context of vacuum
stability.

As a starting point, we briefly recapitulate the main features of the
Standard Model at one-loop level.  In the introductory sections we use
$H$ for the actual Higgs scalar, while $\varphi$ denotes a general
real scalar field which can play the role of the Higgs field $H$ in
our toy model. Once we arrive at our toy model which quantitatively
reproduces the Standard Model, we will again use $H$ for the
corresponding scalar Higgs field.

\subsection{Standard Model running}
\label{sec:sm}

The perturbative approach starts from the usual Higgs potential,
generalized to an effective potential by allowing for a  
dependence on the momentum scale $k$
of all parameters. 
Including higher-dimensional operators it reads
\begin{alignat}{5}
V_\text{eff}(k) 
   = \frac{\mu(k)^2}{2} H^2
   + \sum_{n=2} \frac{\lambda_{2n}(k)}{\highscale^{2n-4}} \; 
     \left( \frac{H^2}{2} \right)^n 
   = \frac{\mu^2(k)}{2} H^2
   + \frac{\lambda_4(k)}{4}H^4 
         + \frac{\lambda_6(k)}{8 \highscale^2} H^6 + \cdots \; ,
\label{eq:potential2}
\end{alignat}
where we denote the scalar doublet as $(H/\sqrt{2},0)=((h+v)/\sqrt{2},0)$, 
with $h$ parametrizing the excitation above the expectation value $v$ and
neglecting the Goldstone modes.  In this form it
is not at all clear whether a slightly negative $\lambda_4$ will lead
to an unstable potential.  This depends on the higher-dimensional
couplings $\lambda_{6,8,...}$ which, if sufficiently large, can
obviously stabilize the Higgs potential for all $k < \Mpl$.\bigskip

Standard power counting shows that long-range observables are
dominated by perturbatively renormalizable operators. Hence, one might
think that higher-dimensional operators can generically be ignored
from the outset.  With the so-defined Standard Model the corresponding
Lagrangian consists of all dimension-4 operators. In that case the
question of stability is usually linked to the sign of $\lambda_4$
defined in Eq.\eqref{eq:potential2}. The beta function for any
coupling $g$ is defined as $\beta_g = d g/d \log k$.  In these
conventions the one-loop renormalization group equations for the Higgs
self-coupling $\lambda_4$, the top Yukawa $y$, and the strong coupling
$g_s$ in the Standard Model read
\begin{alignat}{5}
\beta_{\lambda_4} =  \frac{d\,\lambda_4}{d\,\log k}&=\frac{1}{8\pi^2}
 \left[12\lambda_4^2+6\lambda_4 y^2 - 3y^4
 -\frac{3}{2}\lambda_4 \left( 3g_2^2+g_1^2 \right)
 +\frac{3}{16} \left( 2g_2^4+(g_2^2+g_1^2)^2 \right) \right],
\notag \\
\beta_y =  \frac{d\,y}{d\,\log k}&=\frac{y}{16\pi^2} 
 \left[ \frac{9}{2}y^2-8g_s^2-\frac{9}{4}g_2^2-\frac{17}{12}g_1^2
 \right],
\notag \\
\beta_{g_s} =  \frac{d\,g_s}{d\,\log k}&= - \frac{g_s^3}{16 \pi^2} \;
 \left[ 11 - \frac{2}{3} n_f
 \right] \; .
\label{eq:betas_sm}
\end{alignat}
The top Yukawa coupling is linked to the running top mass by $y=\sqrt{2}m_t/v$,
while the Higgs mass is given by $\mh^2 = 2 \lambda_4 v^2$ plus
contributions from higher-dimensional operators. The couplings or the
related running masses $m_t$ and $\mh$ can be translated into the
corresponding pole or on-shell masses, for example to describe the
kinematics of production and decay processes.  The gauge couplings are
$g_s$ for $SU(3)_c$, $g_2$ for $SU(2)_L$ and $g_1$ for $U(1)_Y$. The
number of fermions contributing to the running of the strong coupling
is $n_f$.  In the Standard Model setup of Eq.\eqref{eq:betas_sm} no
explicit higher-dimensional operators occur. However, if they are
generated by Standard Model particle loops, the effect of, e.g., 
an induced $\lambda_6$ 
coupling
is included in the running of the
renormalizable parameters at a higher loop order. We will return to a
more precise discussion of this point in Sect.~\ref{sec:fixed_point}.

To allow for an easy comparison of our toy model to the Standard Model we
can translate the beta function for the Higgs self-coupling into the
running Higgs mass, namely
\begin{alignat}{5}
\beta_{\mh} 
&= \frac{d\,\mh}{d\,\log k}
=\frac{3}{8  \pi^2 v^2 \mh}
 \left[ \mh^4 
      + 2 \mh^2 m_t^2 
      - 4 m_t^4
      + \text{weak terms} \right] \; .
\label{eq:betas_sm_mass}
\end{alignat}
\bigskip

Let us take a brief look at the essential features that arise from the
renormalization group running given by Eq.\eqref{eq:betas_sm}. 
Toward the
UV, the negative contribution of the 
top Yukawa term drives the
Higgs self-coupling $\lambda_4$ to small values and eventually through
zero.  Ignoring higher-dimensional operators, the Higgs potential
seems to become unstable at large energy scales, before the
stabilizing effect of the weak gauge coupling sets in at very high
scales. This is the usual (meta-)stability issue in the perturbative
setting. We will give a more detailed interpretation in
Sect.~\ref{sec:stability}.

For the quantitative behavior of $\lambda_4$ in Eq.\eqref{eq:betas_sm}
it is important that the top Yukawa coupling $y$ also decreases toward the
ultraviolet under the influence of the strong coupling $\alpha_s$. Indeed, for the
situation we want to investigate, $\lambda_4$ and $y$ both decrease
toward the ultraviolet, and the loop contributions from those
couplings to their running get smaller and smaller.  
At some point the
contributions from the weak couplings and $y$ to the running of
$\lambda_4$ become comparable in size.  When their loop contributions
almost cancel, the running of $\lambda_4$ becomes flat,
$\beta_{\lambda_4}\approx 0$. Finally, in the deep ultraviolet the
weak gauge couplings dominate and turn $\lambda_4$ back to positive
values. This behavior is shown in the upper left panel of
Fig.~\ref{fig:modelflows}~\cite{giudice}.

Following this argument, the self-coupling $\lambda_4$ in the Standard
Model appears to first turn negative and finally become positive
again. From this behavior one could conclude that we live in a
so-called meta-stable vacuum if the Higgs potential, including only
operators to dimension-4  is given by
\begin{alignat}{5}
V(H)
\approx \lambda_4(H)\; \frac{H^4}{4} 
\qquad \text{with} \quad H \gg k_\text{EW} \;,
\label{eq:higgs_potential_pert}
\end{alignat}
at high field values.
Here, it is assumed that the effective potential is well approximated
by identifying the RG scale with the field amplitude,
$\lambda_4(H)\equiv\lambda_4(k=H)$.  This potential features two
minima, our electroweak minimum and a global minimum at very large
field values $H\gg\Mpl$.  The latter occurs far outside the region
where the renormalization group equations of Eq.\eqref{eq:betas_sm}
can be trusted.  We therefore focus on stabilizing effects that set in
below the Planck scale.

\subsection{Toy model}
\label{sec:model}

In this section we set up a simple model that exhibits most of the
essential features of the behavior of the running Standard Model
couplings without inheriting the full gauge structure.  First, we
replace the Higgs field $H$ as part of an $SU(2)$ doublet by a general
real scalar field $\varphi$ featuring a discrete $\mathbb{Z}_2$ chiral
symmetry. This ensures that no Goldstone modes alter the
renormalization group flow in the symmetry-broken regime, just as in
the Standard Model with the full gauge structure.  In a simple Yukawa
system without gauge degrees of freedom we already observe a similar
perturbative flow toward negative $\lambda_4$ at high
scales~\cite{gies2013}. The running of the top Yukawa coupling
to smaller values in the ultraviolet is included when we add an
$SU(3)$ gauge sector. Correspondingly, we investigate the
Euclidean action defined at the UV-cutoff scale $\Lambda$
\begin{equation}
S_{\Lambda} = \int d^4 x 
\left[ \frac{1}{4} F_{\mu\nu}^a F^{a\,\mu\nu} 
      +\frac{1}{2} \left( \partial_\mu \varphi \right)^2 
      + V_\text{eff}(\Lambda)
      + i \sum_{j=1}^{n_f} \overline \psi_j \slashed{D}\psi_j 
      + i \frac{y}{\sqrt{2}} \sum_{j=1}^{n_y} \varphi \, \overline \psi_j \psi_j 
\right] \; ,
\label{eq:GHYaction}
\end{equation}
with an $SU(N_c)$ gauge field ($N_c=3$), a real scalar $\varphi$,
Dirac fermions $\psi_j$ for $n_f$ flavors, the covariant derivative
$D_\mu$ including the strong coupling $g_s$, and the effective scalar
potential $V_\text{eff}$. To avoid confusion, we emphasize that the
Higgs potential $V_\text{eff}(\Lambda)$ is the bare potential from the Standard
Model point of view. Embedding the Standard Model into a more
fundamental theory, the Higgs potential becomes the effective
potential of that theory, containing all effects of quantum
fluctuations above $\Lambda$.  We assume that $n_y < n_f$ of the
fermion species are heavy and shall have a degenerate large Yukawa
coupling $y$. The remaining $n_f-n_y$ flavors have negligible Yukawa
couplings.  The effective scalar potential is now expanded in terms of
dimensionless couplings $\lambda_{2n}$ as
\begin{alignat}{6}
V_\text{eff}(k) =
     \frac{\mu{(k)}^2}{2} \varphi^2
   + \sum_{n=2} \frac{\lambda_{2n}(k)}{k^{2n-4}} \; \left( \frac{\varphi^2}{2} \right)^n \; .
\label{eq:potential_toy2}
\end{alignat}
This potential should be compared to the Standard Model potential
shown in Eq.\eqref{eq:potential2}.  As part of the potential we study
beta functions for the self-interaction $\lambda_4$ and the
$\varphi^6$-coupling $\lambda_6$.  Here we restrict ourselves to the
first in an (infinite) series of possible higher-order
couplings.\bigskip

\begin{figure}[t]
\includegraphics[width=0.48\textwidth]{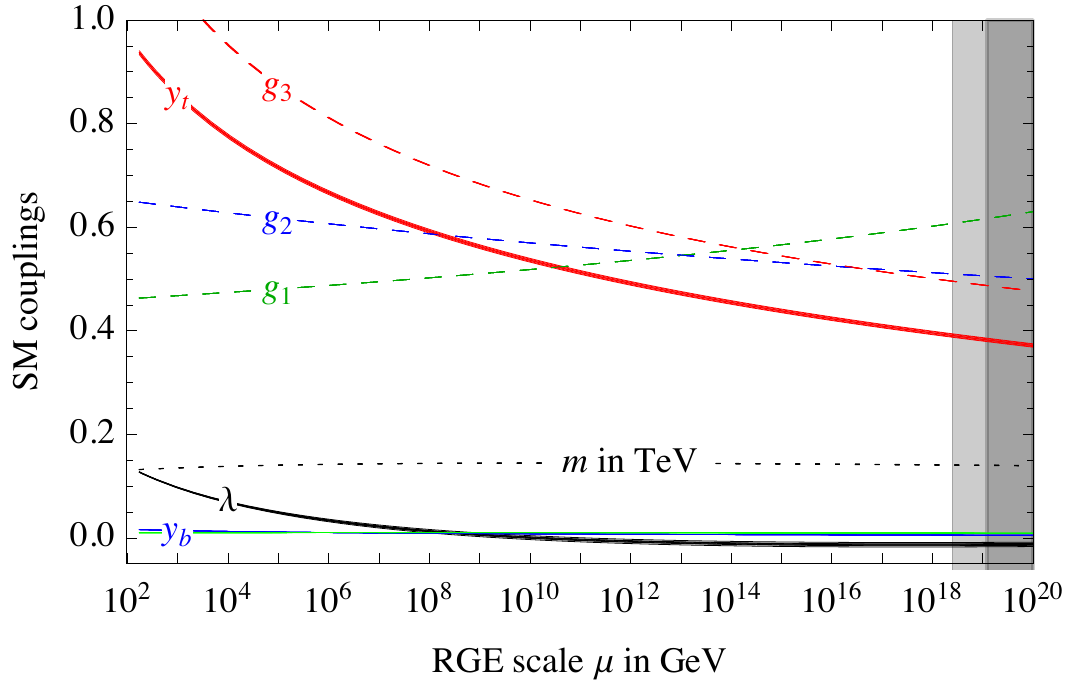}
\hspace*{0.02\textwidth}
\includegraphics[width=0.48\textwidth]{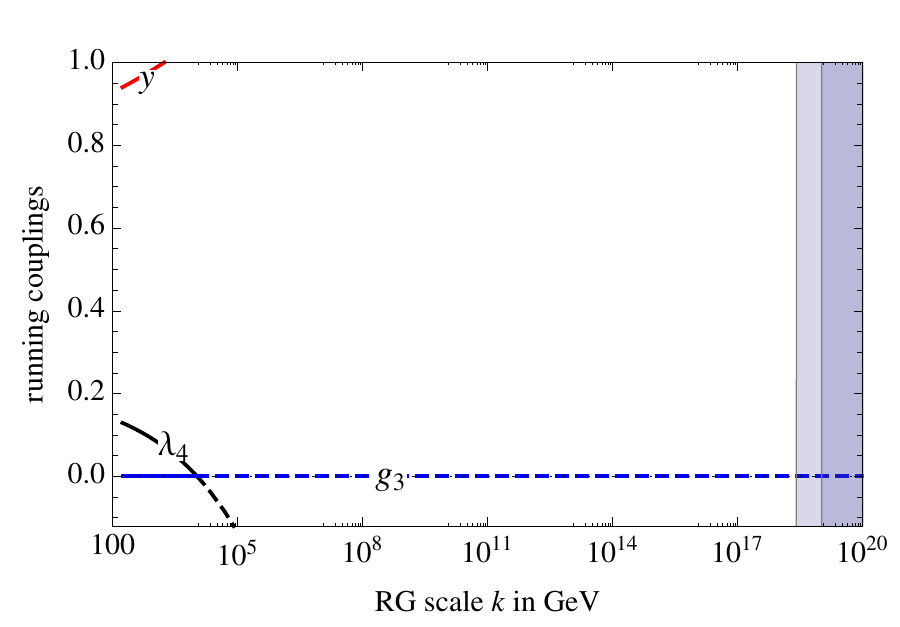} \\
\includegraphics[width=0.48\textwidth]{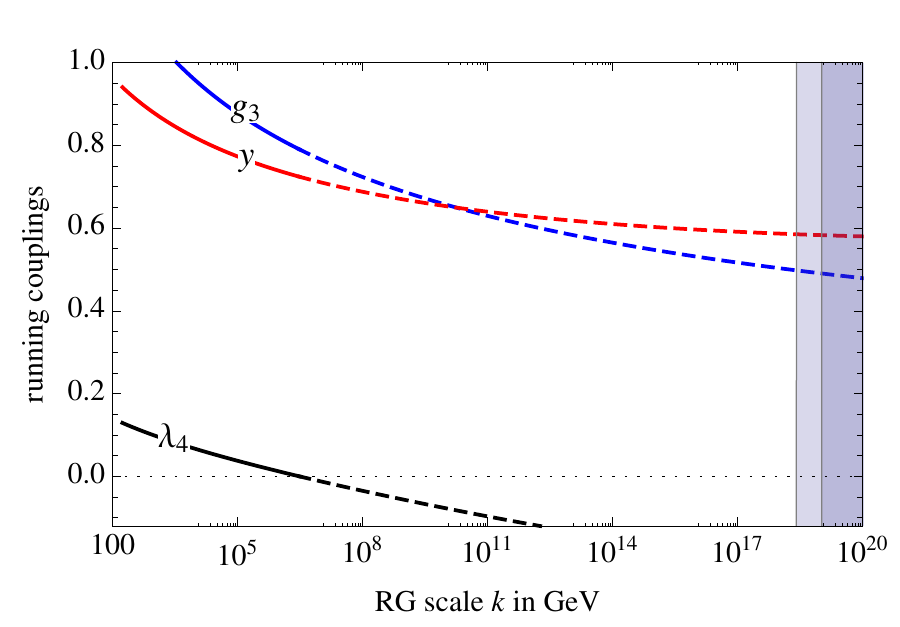}
\hspace*{0.02\textwidth}
\includegraphics[width=0.48\textwidth]{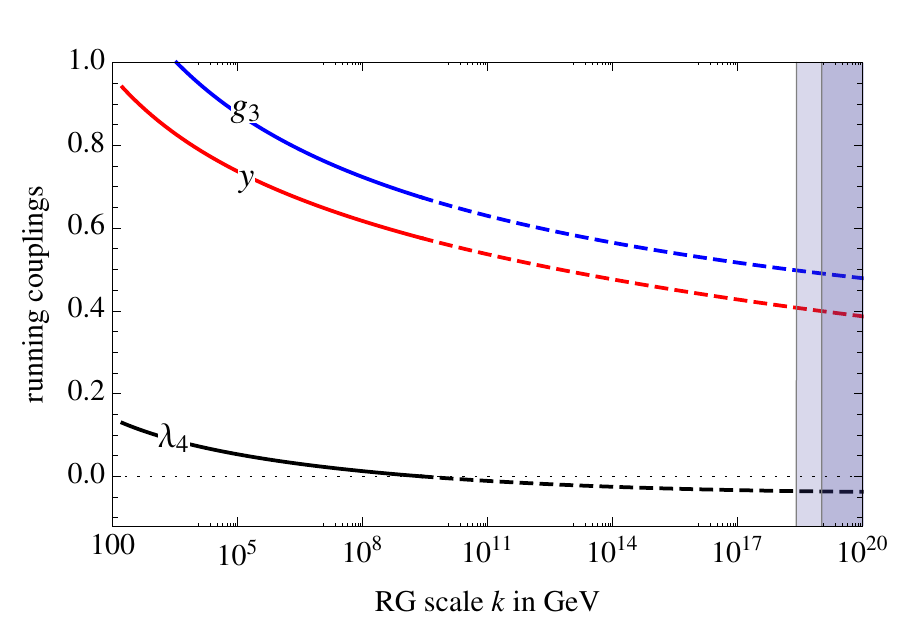}
\caption{Upper left: running of SM couplings, figure taken from
  Ref.~\cite{giudice}.  Upper right to lower right: running
  of our toy model couplings after including the running strong
  coupling and electroweak coupling effects. Dashed lines indicate the
  regime where $\lambda_4(k)<0$, which in the perturbative approach
  defines the loss of vacuum stability.}
\label{fig:modelflows}
\end{figure}

Because our toy model does not reflect the weak gauge structure of the
Standard Model, the gauge couplings $g_1$ and $g_2$ do not
appear. However, we know that in the Standard Model they have
significant effects: first, the gauge couplings give a significant
positive contribution to $\beta_{\lambda_4}$, balancing the negative
top Yukawa terms for small values of $\lambda_4$; second, they decrease the Yukawa coupling in the UV, thereby also increasing
$\lambda_4$ toward large scales. Since the variation of the weak
coupling at large energy scales is modest, we account for its effects
by including a finite contribution in the beta functions for
$\lambda_4$ and $y$, parametrized by a fiducial coupling $g_F$ and
numerical constants $c_\lambda, c_y$.
\begin{alignat}{5}
\beta_{\lambda_4} &= \frac{1}{8\pi^2} \;
 \left[ - n_y N_c y^4+2 n_y N_c y^2\lambda_4 
        + 9\lambda_4^2 
        - \frac{15}{4} \lambda_6 
        + c_\lambda g_F^4
 \right] \notag \\
\beta_{\lambda_6} &= 2\lambda_6 +\frac{1}{16\pi^2}
 \left[ 2 n_y N_c y^6
      + 6 n_y N_c y^2\lambda_6 
      - 108\lambda_4^3
      + 90\lambda_4 \lambda_6 
 \right] \notag\\
\beta_y &= \frac{y}{16\pi^2} \; 
  \left[ \frac{3+2n_y N_c}{2} y^2 
        - 3 \frac{N_c^2-1}{N_c} g_s^2 
        - c_y g_F^2 
  \right] \notag \\
\beta_{g_s} &= - \frac{g_s^3}{16\pi^2} \; 
  \left[ \frac{11}{3}N_c-\frac{2}{3}n_f \right] \; .
\label{eq:betas_toy}
\end{alignat}
The scalar mass is given by $m_\varphi^2 = 2 \lambda_4 v^2$, and the running
top mass by $y=\sqrt{2}m_t/v$. To approximate the Standard Model we
choose $n_f=6$ to account for the contribution of all flavors to the
running of the strong gauge coupling. As long as we only keep the top quark
contribution to the running of the Higgs quartic coupling, we set
$n_y=1$. The expressions in Eq.\eqref{eq:betas_toy} reproduce the
standard one-loop beta functions for our model if higher-dimensional
operators are ignored~\cite{Harada:1994wy}.

We have verified that the simple modelling of the electroweak
  gauge boson effects in Eq.~\eqref{eq:betas_toy} already provides a
  reasonable estimate. In a more self-consistent treatment,
  electroweak contributions to all running couplings $\mu^2,\lambda_4,\lambda_6\dots$
  of the Higgs potential appear. This is discussed in more detail in
  Appendix~\ref{advancedfudge}.\bigskip

The running of the strong coupling and the top Yukawa coupling in our model
agree exactly with that of the Standard Model given in
Eq.\eqref{eq:betas_sm}.  Only the running of the scalar quartic vs
Higgs quartic coupling is slightly different because of different numbers of degrees of
freedom.  To compare our toy model with the Standard Model we also
write the beta function for the scalar quartic coupling in terms of
the scalar mass as in Eq.\eqref{eq:betas_sm_mass},
\begin{alignat}{5}
\beta_{m_\varphi} 
&= \frac{3}{8\pi^2 v^2 m_\varphi} \;
 \left[ \frac{3}{4} m_\varphi^4
       +2 m_t^2 m_\varphi^2
       -4 m_t^4 + \text{weak and $\lambda_6$ terms} \right] \notag \\
&= \frac{3}{8\pi^2 v^2 m_\varphi} \;
 \left[ (0.93  m_\varphi)^4
       +2 m_t^2 m_\varphi^2
       -4 m_t^4 + \text{weak and $\lambda_6$ terms} \right] \; .
\label{eq:betas_toy_mass}
\end{alignat}
The only difference is the self-interaction contribution to the
running of the scalar or Higgs mass. This means that we can
model the running of the top--Higgs sector if we slightly shift
the scalar and top mass ratio at the per-cent level. \bigskip

In Fig.\ref{fig:modelflows} we illustrate our toy model with
representative one-loop renormalization group flows in three steps:
\begin{enumerate} 
\item in the upper left panel we show the full 3-loop Standard Model
  running from Ref.~\cite{giudice} as reference;
\item in the upper right panel we include a scalar field and one heavy
  (top) quark, $n_y=1$, without any contributions from the gauge
  couplings (pure $\mathbb{Z}_2$ model);
\item in the lower left panel we add the effect of $\alpha_s$ on the
  running top Yukawa coupling; the strong coupling itself 
  runs with
  $n_f=6$ active flavors;
\item finally, in the lower right panel we also include a constant
  fiducial coupling with $c_\lambda=9/16, c_y = 97/30$ and $g_F=0.57$ to model the effect of the weak coupling.
\end{enumerate}
Since this choice of parameters allows us to quantitatively reproduce
the behavior of the Higgs sector in the Standard Model, we refer to
the corresponding scalar field given the appropriate model
parameters as the `Higgs scalar', \ie $\varphi \to H$. For the flow
trajectories we choose the running top mass such that
$m_t(m_t)=164~\gev$~\cite{giudice}, and correspondingly the
Higgs mass at $\mh=2\lambda_4(m_t)v^2 =125~\gev$. This means that we
ignore the small difference between the running Higgs mass $\mh(m_t)$
and the measured Higgs pole mass. The running of the strong coupling
satisfies $\alpha_S(m_Z)=0.1184$ at
$m_Z=91~\gev$~\cite{giudice}.

When we include the strong coupling in the bottom left panel of
Fig.~\ref{fig:modelflows} the top Yukawa coupling decreases toward
the ultraviolet. This allows for lower Higgs mass bounds from vacuum
stability compared to the pure $\mathbb{Z}_2$ model~\cite{gies2013}.  With the
additional approximate effects from the weak coupling we also
reproduce the back-bending toward positive $\lambda_4$ in the deep
ultraviolet, as it occurs in the Standard Model, shown in the bottom
right panel of Fig.~\ref{fig:modelflows}. The weak coupling term $c_y$
in the flow of the top Yukawa coupling is responsible for the faster decrease
of the Yukawa coupling, such that it flows almost parallel to the strong gauge
coupling. Thereby also the flow of $\lambda_4$ becomes flatter. The
weak coupling contribution $c_\lambda$ to the running of the Higgs
quartic coupling has a smaller impact at intermediate scales and
induces the final back-bending of $\lambda_4$ close to the Planck
scale.

\subsection{The issue of vacuum stability}
\label{sec:stability}

In this section we clarify some aspects of the relation between the
perturbative and non-perturbative approaches. First we address a
potential confusion whether the top loop stabilizes or de-stabilizes
the potential. Second, in the perturbative approach the RG scale is
replaced by the field value. In the non-perturbative approach both,
the RG scale and the field value appear explicitly. Finally, in the
perturbative approach one usually employs a massless regulator based
on dimensional regularization. The question arises what happens in the
presence of a finite, physical UV-cutoff.\bigskip

Let us start by briefly recalling how the issue of vacuum stability
arises and what role a finite cutoff plays.  We refer to a potential
as stable if it is bounded from below, otherwise it is unstable.  A
stable potential can exhibit several minima, where a local minimum
might be meta-stable while the global minimum is stable.  As argued
above, the top loop is by far the dominant factor in the
renormalization group evolution of the Higgs potential. Indeed, the
whole issue of stability can already be understood qualitatively by
considering the term $\beta_{\lambda_4} = -3y^4/(8\pi^2) + \cdots$ in
Eqs.\eqref{eq:betas_sm} and~\eqref{eq:betas_toy}. Integrating the
$\beta$-function toward the ultraviolet, this term drives the quartic
self-coupling to negative values at some high scale. This is the usual
argument for the loss of stability in the Higgs potential if the Higgs
potential is approximated by $V \approx \lambda_4(H)H^4/4$.

This observation appears to be in conflict with non-perturbative
lattice simulations~\cite{Holland:2003jr} and with arguments that the
interacting part of the top loop is non-negative and cannot
induce any instability in the presence of a finite UV-cutoff scale
$\Lambda$~\cite{Gies:2014xha,Branchina:2005tu}. To understand this in
more detail, let us consider a one-loop calculation of the effective
potential at the electroweak scale $k\approx k_\text{EW}\approx 0$,
where we neglect $k_\text{EW}\ll \Lambda$ when appropriate. As shown
in Ref.~\cite{Gies:2014xha} the fermion determinant representing the
top contribution to the effective potential for any finite cutoff
$\Lambda$ can be written as
\begin{equation}
\Delta V_\text{top} =-c_2 \Lambda^2 H^2+ \text{positive terms,}
\label{eq:deltav1}
\end{equation}
where $c_2>0$.  We start from a stable bare potential $V_\text{UV} =
V_\text{eff}(\Lambda)$ of the kind shown in Eq.\eqref{eq:potential2}
with quadratic and quartic Higgs terms. The top loop then gives us an
approximate weak-scale potential
\begin{equation}
V_\text{eff} (k_\text{EW})\approx 
V_\text{UV} +\Delta V_\text{top} =
\frac{\mu^2(\Lambda) -c_2 \Lambda^2}{2} H^2+\frac{\lambda_4\left(\Lambda\right)}{4}H^4 
+\text{positive terms,}
\label{eq:veff_stable1}
\end{equation}
and will never lead to an unstable effective potential at the weak
scale. A closer look at the contribution from the top loop below the
cutoff scale $\Lambda$ reveals the formally sub-leading terms 
\begin{alignat}{5}
\Delta V_\text{top} 
&= -\frac{1}{4\pi^2}\int^{\Lambda}_{0}dq\, q^3\log\left(1+\frac{y^2H^2}{2q^2}\right)
\notag \\
&=-\frac{y^2}{16\pi^2}\Lambda^2H^2+\frac{1}{64\pi^2}\left[y^4H^4\log\left(1+\frac{2\Lambda^2}{y^2H^2}\right)+2y^2\Lambda^2H^2-4\Lambda^4\log\left(1+\frac{y^2H^2}{2\Lambda^2}\right)\right] \notag \\
&=-c_2 \Lambda^2 H^2+ c_4 \frac{y^4}{4} H^4 \; \log \frac{\Lambda}{k_\text{EW}} +\ldots \; .
\label{eq:deltav2}
\end{alignat}
where we normalize the contribution such that $\Delta
V_\text{top}(H=0)=0$.  In the last line we adopt the general
notation of Eq.\eqref{eq:deltav1}.  The additional logarithmic
contribution to the quartic potential term contributes to the
Higgs mass as~\cite{branchina,gies2013}
\begin{equation}
\lambda_4(k_\text{EW})
= \lambda_4(\Lambda)+ c_4 y^4\log \frac{\Lambda}{k_\text{EW}} \approx \frac{1}{8}\; .
\label{eq:cutoffrunning}
\end{equation}
This is nothing but the renormalization of the Higgs self-coupling.
For a sufficiently large cutoff scale $\Lambda$ the above top
contribution forces us to choose a negative high-scale value
$\lambda_4(\Lambda)<0$ to obtain the measured Higgs mass.

While no instability in the effective potential is induced by the top
loop, its contribution is such that we would have to choose an
unstable potential at the ultraviolet scale $\Lambda$ to reproduce the
measured Higgs mass in the infrared. The usual perturbative
renormalization group evolution simply determines the value we would
have to choose in the ultraviolet in the class of pure
$H^4$-potentials and in that sense reflects the fact that we would
have to choose an unstable UV-potential if we want to start at a very
high scale $\Lambda$.\bigskip

\begin{figure}[b!]
\includegraphics[width=0.28\linewidth]{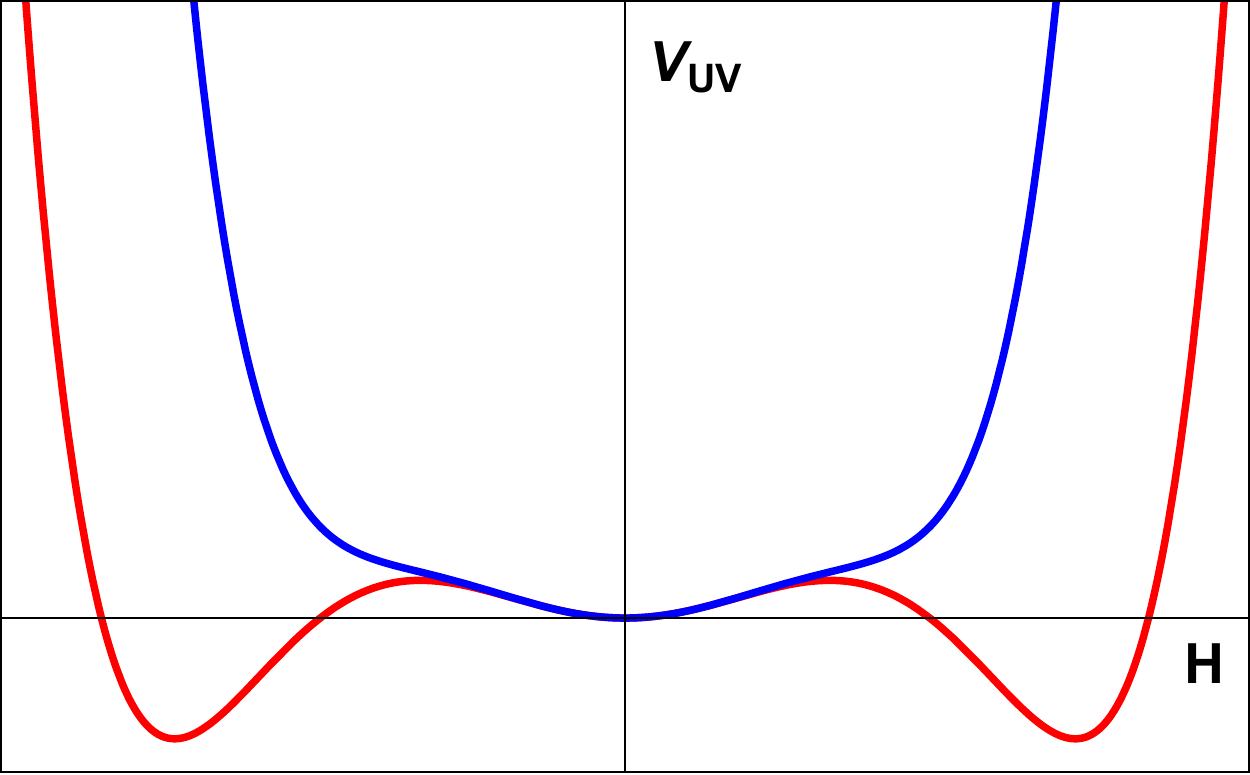}
\hspace*{0.9cm}
\includegraphics[width=0.28\linewidth]{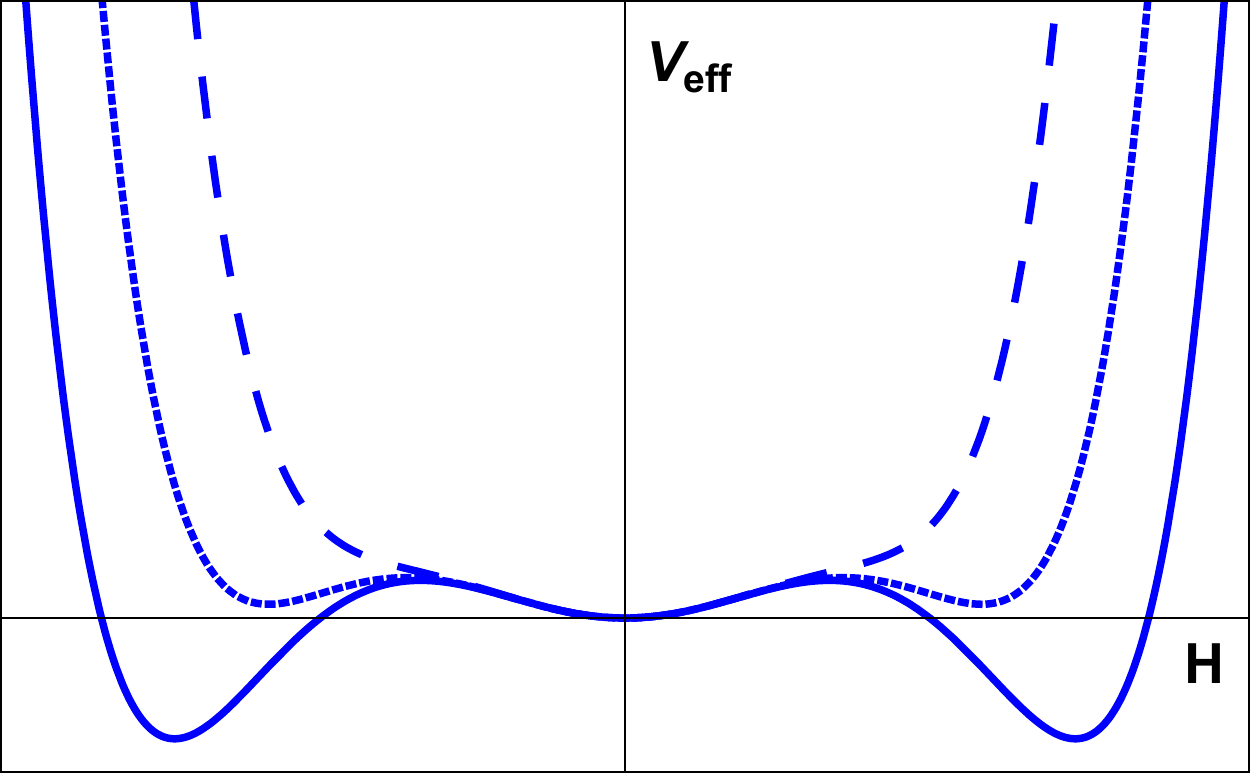}
\caption{Left: Sketch of the possible shape of $V_\text{UV}$ with $\mu^2(\Lambda)>0$
  but $\lambda_4(\Lambda)<0$. 
  Potentials are stabilized by $\lambda_6(\Lambda)$ smaller (red) and larger (blue)
  than $\lambda_4^2(\Lambda) \Lambda^2/(3 \mu^2(\Lambda))$, respectively. 
  Right: Sketch of different possible shapes of the effective potential $V_{\rm eff}(k\approx 0)$ that may occur depending  on the size of
  $\lambda_6(\Lambda)$, all corresponding to a stable UV-potential
  with a single minimum at $H\approx 0$ as the blue curve in the left
  panel. Note that the resolution of this sketch is not high enough to display the offset of the electroweak vacuum from $H=0$, the corresponding two minima appear as one at $H=0$.
}
\label{fig:potentialUVEW2}
\end{figure}

\begin{figure}[t]
\includegraphics[width=0.65\linewidth]{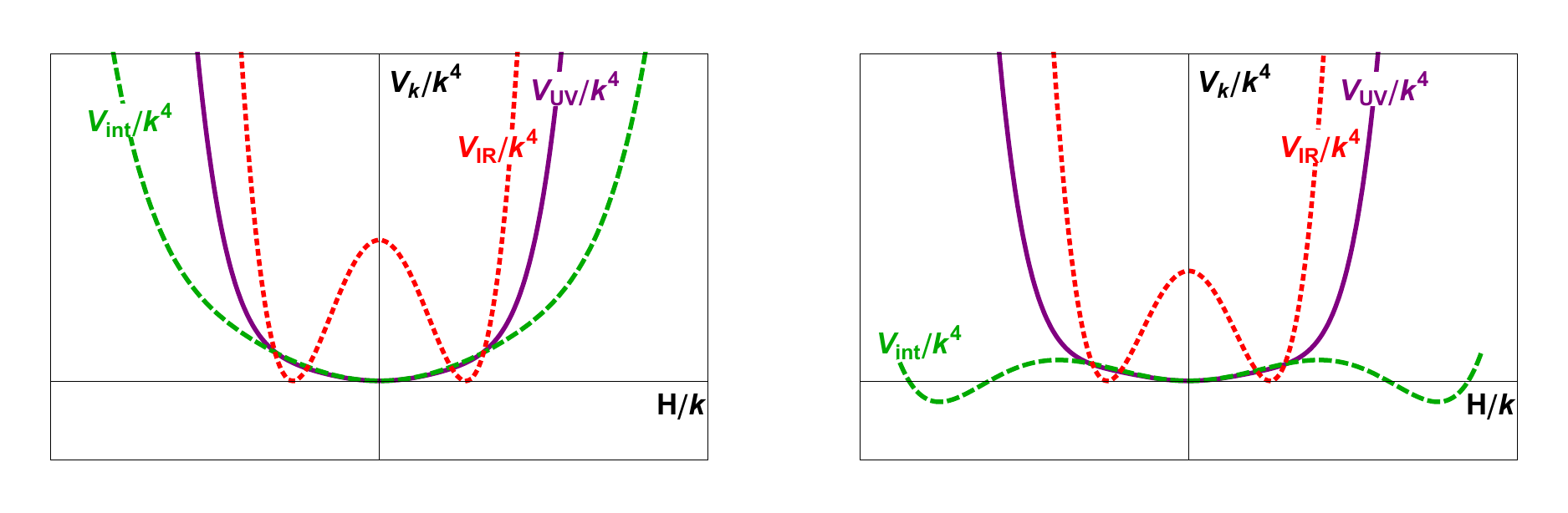}
\caption{Starting from a stable potential in the UV different flows to the IR potential with the electroweak minimum are possible. In the left
panel we show a situation where the potential remains stable during the whole flow from $k=\Lambda$ to the point where
the electroweak minimum forms. This corresponds to region I in Fig.~\ref{fig:lambda6cutoff}.
For smaller initial values of $\lambda_{6}$ corresponding to region II in Fig.~\ref{fig:lambda6cutoff}, we observe a meta-stable behavior of the scale dependent potential already at intermediate values of $k$, which then disappears. The electroweak symmetry breaking minima are not formed until later.
We call this behavior pseudo-stable.}
\label{fig:potentialUVEW1}
\end{figure}

Stabilizing the UV-potential requires higher-dimensional
operators. In this situation the stability of the UV-potential cannot
be determined from the running of $\lambda_4$ alone, but requires the
full UV-potential. As a simple example, we include a $\lambda_6$ term
as in Eq.\eqref{eq:potential2},
\begin{equation}
V_\text{UV} = V_{\rm eff}(\Lambda)
= \frac{\mu^2(\Lambda)}{2} H^2
 +\frac{\lambda_4(\Lambda)}{4}H^4
 +\frac{\lambda_6(\Lambda)}{8\Lambda^2}H^6\; .
\label{eq:stability_pot6}
\end{equation}
As we have seen, $\lambda_4(\Lambda)$ is essentially fixed by the
measured value of the Higgs mass in the infrared. For sufficiently
large $\Lambda$ it is negative. Similarly $\mu^2(\Lambda)$ is fixed by
requiring $v=246~\gev$ in the infrared. For the simple momentum-space
cutoff regularization, it is typically large and positive,
$\mu^2(\Lambda)\sim\mathcal{O}(0.01)\Lambda^2$.  As expected,
$\lambda_6(\Lambda)$ is the only new and free parameter. Its value is
essentially undetermined by measurements in the infrared, because it
is suppressed by the large scale $\Lambda$.

Choosing $\lambda_6(\Lambda)$ positive ensures that the UV-potential
grows at large field values, making it a viable bare potential for a
quantum field theory. In principle, this can be achieved with an
arbitrarily small value of $\lambda_6(\Lambda)$. Since
$\mu^2(\Lambda)>0$, the potential then has a qualitatively similar form
to the red curve in the left panel of Fig.~\ref{fig:potentialUVEW2}
with a local minimum at $H=0$ and a global minimum at a large field
value $H\neq 0$.  Choosing
\begin{equation}
\lambda_6(\Lambda)> \lambda_4^2(\Lambda) \; \frac{\Lambda^2}{3\mu^2(\Lambda)}
\label{eq:stability}
\end{equation}
removes the second minimum at $H\neq 0$ for negative
$\lambda_4(\Lambda)$. The only minimum appears at $H=0$, as shown by
the blue curve in Fig.~\ref{fig:potentialUVEW2}.  In this paper we are mainly
interested in this type of potentials.\bigskip

Nevertheless, a stable UV potential with a single minimum at $H=0$
according to Eq.\eqref{eq:stability} does not necessarily imply that
the effective potential only features a single minimum.  In the course
of the RG flow, all types of behavior with a potential bounded from
below shown in Fig.~\ref{fig:potentialUVEW1} do
occur, depending on the choice of UV parameters.

Qualitatively, this behavior can be understood from the unique-vacuum
condition in Eq.\eqref{eq:stability}, which can be applied to any
scale $k$.  
While $\lambda_6(k)$ decreases rapidly for decreasing $k$, 
 $\lambda_4$ grows at the same time and ultimately becomes
positive.  If $\lambda_{4}$ turns positive first, then the scale
dependent potential is always stable during the RG evolution in our
approximation, cf. left panel in Fig.~\ref{fig:potentialUVEW1}.  On the other hand, if Eq.\eqref{eq:stability} is
violated first, a second minimum appears and the minimum at $H=0$ can
become meta-stable for some range of RG scales, cf. right panel in Fig.~\ref{fig:potentialUVEW1}. However, a reliable
(meta-)stability analysis of the full potential will require to go
beyond the simple polynomial expansion of the effective potential
around one minimum.

For completeness let us note what happens when we continue to run to
lower energy scales. Independently of the high-scale behavior
discussed above, $\mu^2(k)$ drops below zero near the electroweak
scale. For these low scales $\lambda_{4}$ is always positive and the
interplay with the negative $\mu^2(k)$ results in a finite vacuum
expectation value $v$. In the presence of higher-dimensional
operators, we rather generically find RG flows that interconnect a
stable UV-potential with a minimum at $H=0$ with stable effective
potentials in the infrared and a global electroweak minimum at
$H=v$. If the flow passes through a finite regime where our polynomial
approximation looks meta-stable, we refer to the scenario as
pseudo-stable. In that case a full stability analysis requires a detailed
non-perturbative analysis of the effective potential.\bigskip

Our discussion shows that we have to distinguish between the stability
of the UV-potential where no quantum fluctuations below $\Lambda$ are
taken into account and that of the effective potential with all fluctuations included. In principle the UV potential and the effective potential can have quite different shapes as sketched in the left and right panels of Fig.~\ref{fig:potentialUVEW2}. In the
approximation $V\sim \lambda_4(H)H^4/4$, this difference cannot be
accounted for properly: on the one hand the running quartic coupling
defines the UV-potential as $V(\Lambda) = \lambda_4(\Lambda) H^4/4$.
On the other hand, the identification $k \rightarrow H$ is assumed to
be a good approximation for the effective potential. This means that
one and the same function $\lambda_4(k)$ determines the stability of the
UV-potential and the stability of the effective potential. The
presence of higher-dimensional operators 
influences both
aspects: at the UV-scale, they can modify the UV-potential in a rather
general way. Successively, they can contribute to the RG flow of the
renormalizable operators for some range of scales.  The functional
renormalization group takes these aspects into account by describing a
scale-dependent effective potential $V_\text{eff}(k;H)$, depending on
both the scale $k$ as well as the field value, such that
\begin{equation}
V(k=\Lambda;H)=V_\text{UV}
\qquad \text{and} \qquad 
V(k=0;H)=V_\text{eff}.
\end{equation}
In this manner, the (meta-)stability properties of the effective
potential can be followed in a scale-dependent manner.\bigskip

Finally, let us note another subtlety in the comparison between the
perturbative and non-perturbative approaches. The perturbative
approach usually relies on a mass-independent regularization scheme.
To avoid large threshold effects one should therefore stop the running
of the couplings at the appropriate mass scales.  Most of the relevant
masses are proportional to 
$H$.  This suggests to approximate
\begin{equation}
V_\text{eff}=V(k=H,H)\approx \frac{1}{4}\lambda_{4}(H)\,H^4 \; ,
\end{equation}
as is usually done in the perturbative approach.  In
Eq.\eqref{eq:cutoffrunning} we have already identified the running
Higgs self-coupling at the scale $k_\text{EW}$.  Using the
identification $k=H$ we essentially reproduce 
the first term in the
square brackets of the second line of Eq.\eqref{eq:deltav2}.  Let us
check this explicitly,
\begin{alignat}{5}
\label{potexpansion}
V_\text{eff} (k_\text{EW})&\approx 
V_\text{UV} +\Delta V_\text{top} 
\notag \\
&=
\frac{\mu^2(\Lambda)
-c_2 \Lambda^2}{2} H^2+\frac{1}{64\pi^2}\left[y^4H^4\log\left(1+\frac{2\Lambda^2}{y^2H^2}\right)+2y^2\Lambda^2H^2-4\Lambda^4\log\left(1+\frac{y^2H^2}{2\Lambda^2}\right)\right]+\frac{\lambda_{4}(\Lambda)}{4} H^4
\notag \\
&=
\frac{\mu^2
(k_\text{EW})}{2} H^2+\frac{1}{64\pi^2}\left[y^4H^4\log\left(1+\frac{2\Lambda^2}{y^2H^2}\right)+2y^2\Lambda^2H^2-4\Lambda^4\log\left(1+\frac{y^2H^2}{2\Lambda^2}\right)\right]+\frac{\lambda_{4}(\Lambda)}{4}H^4
\notag \\
&=
\frac{\mu^2(k_\text{EW})}{2} H^2
+\frac{1}{4}\left[\lambda_{4}(\Lambda)+\frac{y^4}{8\pi^2}\log\left(\frac{\Lambda}{H}\right)+\frac{y^4}{32\pi^2}-\frac{y^4}{16\pi^2}\log\left(\frac{y^2}{2}\right)\right]H^4+{\mathcal{O}}\left(H^4\frac{H^2}{\Lambda^2}\right)
\notag \\
&=
\frac{1}{4}\lambda_{4}(k=H)H^4+\frac{1}{8\pi^2}({\text{finite\,\,in}\,\,}\Lambda)H^4+{\mathcal{O}}\left(H^4\frac{H^2}{\Lambda^2}\right)+{\mathcal{O}}\left(H^4\frac{\mu^2
(k_\text{EW})}{H^2}\right).
\end{alignat}
However, $V_\text{eff}=\lambda_{4}(H)H^4/4$ only holds for $H\gg v\sim
|\mu(k_\text{EW})|$ and $\Lambda\gg H$. This means that in the
presence of a finite, physical UV-cutoff $\Lambda$ the perturbative
running of $\lambda_{4}(H)$ provides a good approximation of the
effective potential only for field values $H\ll \Lambda$.  Indeed,
using the perturbative running for a cutoff-free mass-independent
regularization scheme can make the effective potential appear to be
unstable beyond $H\sim\Lambda$ even when it is
not~\cite{Holland:2003jr,Gies:2014xha}. Essentially, this results from
a breakdown of the expansion in powers of $H/\Lambda$ when going from
the second to the third line in Eq.\eqref{potexpansion}.  For the use
of the perturbative approach we therefore always have the condition
that the field values --- more generally any scale under consideration
--- are smaller than any physical cutoff.

\subsection{Higgs mass bounds}
\label{sec:higgsmass}

\begin{figure}[t]
\includegraphics[width=0.45\linewidth]{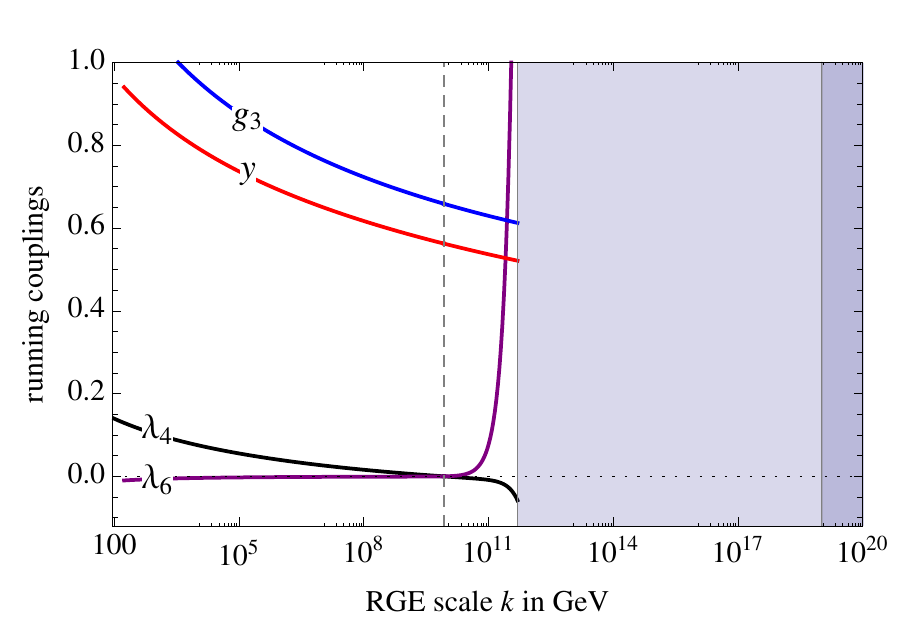}
\includegraphics[width=0.45\linewidth]{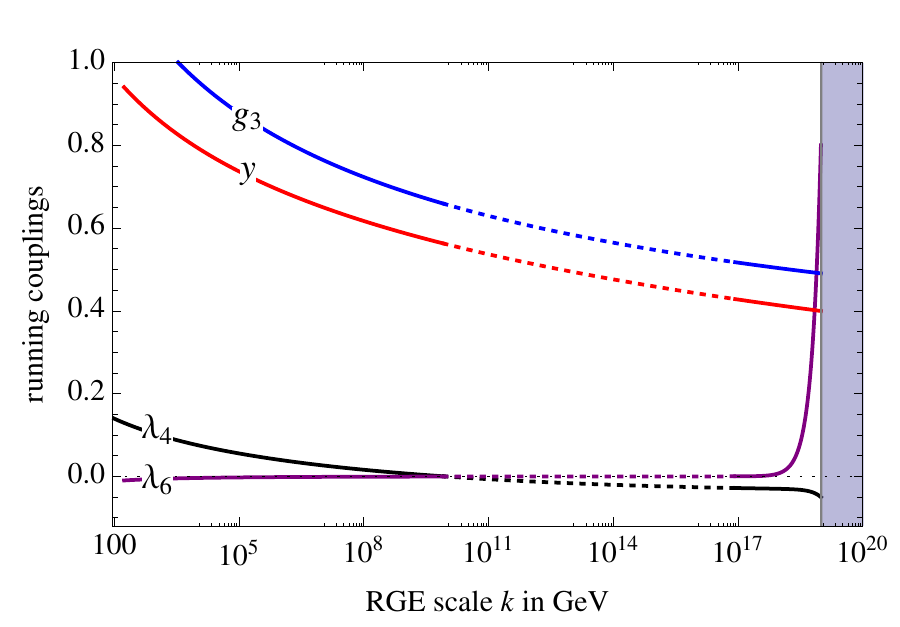}
\caption{Running couplings including $\lambda_4$ and $\lambda_6$,
  corresponding to Eq.\eqref{eq:betas_toy} with $\mh = 125~\gev$.  The
  bare potentials satisfy the stability condition of
  Eq.\eqref{eq:stability} and thus remain bound from below at all
  lower scales.  Left: $\lambda_4(\Lambda)=-0.058$ and
  $\lambda_6(\Lambda)=2.0$ and $\Lambda=5\cdot10^{11}~\gev$. Right:
  $\lambda_4(\Lambda)=-0.050$ and $\lambda_6(\Lambda)=0.8$ and
  $\Lambda=10^{19}~\gev$.}
\label{fig:LambdaLambdap}
\end{figure}

When it comes to stability considerations, the main novel feature of
our renormalization group flows is the presence of a second model parameter,
$\lambda_6$, as a representative of general higher-dimensional operators. 
In the absence of $\lambda_6$ the stability condition for the bare potential,
$\lambda_4(\Lambda)=0,$ at the so-defined ultraviolet cutoff scale
$\Lambda$ constitutes the lowest viable choice for the quartic
coupling. 
Positive higher-dimensional couplings in the ultraviolet
can allow for negative values of the Higgs quartic coupling and yet a stable
potential. A negative quartic coupling in the ultraviolet also lowers
the infrared quartic coupling value, leading to a stable potential with
a Higgs mass below the conventional lower bound obtained when neglecting
$\lambda_6$.

Let us now consider two examples for the influence of $\lambda_{6}>0$
as shown in Fig.~\ref{fig:LambdaLambdap}.  In the first case,
corresponding to the left panel of the figure we assume a relatively
large UV-value of $\lambda_{6}$.  At first, the effect of this term is
sizable and therefore the RG flow deviates significantly from the
usual perturbative scenario.  However the canonical dimension of
$\lambda_6$ implies that it quickly decreases toward the infrared,
while $\lambda_4$ increases. At some scale the combined
renormalization group flow reaches the point $\lambda_6 \approx 0$ and
$\lambda_4 \approx 0$. This setup coincides with the ultraviolet
starting point of the perturbative approach neglecting $\lambda_6$ and
marks the boundary of stability.  The infrared regime to the left of
the dashed line in the left panel, where $\lambda_4$ again reaches
positive values, is basically identical to the usual perturbative
scenario. This means that due to higher-dimensional couplings our
scenario features a significantly increased cutoff scale, while
reaching the same values for the Higgs mass, quartic self-coupling,
and top mass in the infrared. The naive upper limit on the validity of
the model near $10^{10}~\gev$ (dashed line) is replaced by a higher
scale $\sim 10^{12}~\gev$, at the expense of a nontrivial running of
higher-dimensional operators. In between these two scales an effective
field theory description within the model is still possible, while
true new physics would presumably have to set in beyond
$10^{12}~\gev$, in order to control the strong running of the scalar
couplings.
 
In the second scenario, shown in the right panel of
Fig.~\ref{fig:LambdaLambdap}, we choose a smaller value of
$\lambda_{6}>0$ and a much larger cutoff scale $\Lambda$. In spite of
this, the UV-potential at $\Lambda$ is still stable with a
global minimum at $H=0$.  As we have chosen a significantly larger
cutoff scale $\Lambda$, $\lambda_{6}$ becomes smaller than required by
the unique-vacuum condition Eq.\eqref{eq:stability} before
$\lambda_{4}$ reaches positive values. This implies an intermediate pseudo-stable region.
The existence of this intermediate region
indicated by dashed lines is a characteristic of the
pseudo-stable scenario mentioned above. In our simple polynomial
expansion, the scale-dependent effective potential develops a second
minimum and $H=0$ seems to be only a meta-stable minimum for
intermediate RG scales.
Our analysis based on the functional RG as outlined in
App.~\ref{app:compare} going beyond the simple perturbative arguments
indicates that this behavior should be assessed critically.  We
tentatively consider it an artifact of the polynomial expansion of the
effective potential around one minimum. The functional RG provides us
with a $\beta$-function for the whole potential,
Eq.\eqref{eq:potflow}. We evaluate it in a polynomial expansion about
an expansion point $H=0$ at high scales and $H=v$ near the weak
scale. After integrating the flow we check whether the solution for
the scale-dependent effective potential satisfies this
$\beta$-function. Whereas the accuracy of the solution is indeed at
the level of machine precision near the expansion point, the accuracy
depletes away from this point. The convergence radius of this
expansion is typically  
of
the order of $\mh$~\cite{gies2013}. In
particular, near the apparent new minima the polynomial expansion
should not be trusted. Reliable statements about the stability of the
effective potential in pseudo-stable scenarios require a dedicated
analysis of the RG flow of the effective potential.

\begin{figure}[t]
\includegraphics[width=0.5\linewidth]{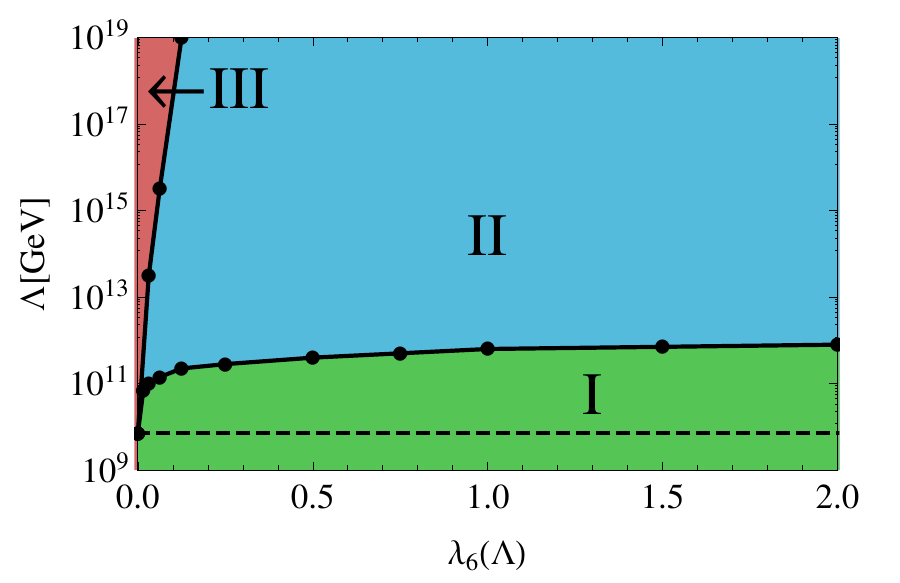}
\caption{Different stability regions as a function of $\Lambda$ and
  $\lambda_{6}$. In region~I 
  (green)
  the potential is stable
  everywhere; in region~II  
  (blue)
  the UV-potential is stable, while the potential is only pseudo-stable for
  intermediate scales $v < k < \Lambda$; in region~III
  (red)
  the UV-potential violates the unique-minimum condition.
  Our simple model includes the effects of the electroweak gauge bosons only in the running of $\lambda_{4}$. In Appendix~\ref{advancedfudge}, we model the electroweak contributions to the full Higgs potential. The corresponding region III is then somewhat larger, cf. Fig.~\ref{fig:lambda6cutoff:adv}.}
\label{fig:lambda6cutoff}
\end{figure}

An improved functional RG analysis according to App.~\ref{app:compare} also
clarifies the following apparent problem: in the infrared, the
$\lambda_6$ coupling runs negative seemingly indicating a global
instability. This is prevented by even higher-operators $\sim H^8$,
which are taken into account in our improved functional RG analysis that we employ to determine the stability regions shown in Fig.~\ref{fig:lambda6cutoff}
discussed below. All our explicit solutions for the scale-dependent effective
potential are bounded from below.
\bigskip

\begin{figure}[t]
\includegraphics[width=0.48\linewidth]{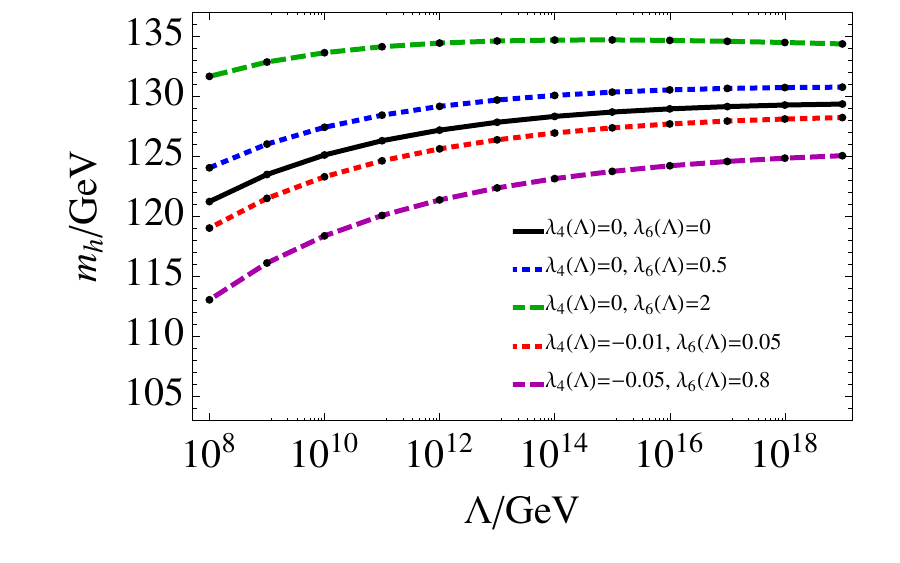}
\includegraphics[width=0.48\linewidth]{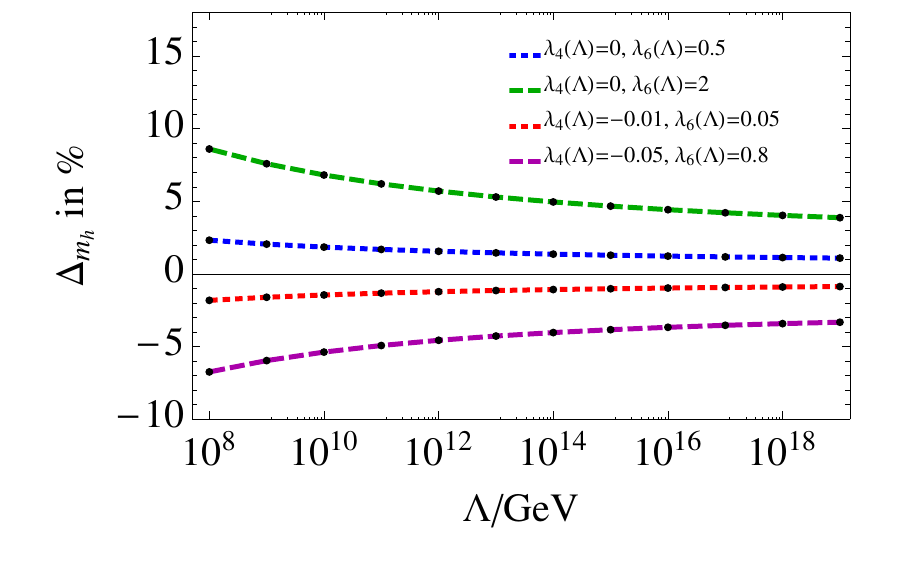}
\caption{\label{fig:deltamhplot} Infrared value of the Higgs mass
  (left) and the relative shift $\Delta_{\mh}$ measuring the departure
  from the conventional lower bound (right), as a function of the
  UV-cutoff $\Lambda$. We show different UV boundary conditions as
  indicated in the plot.}
\end{figure}

Let us now examine the effects of $\lambda_{6}$ on the stability
bounds for the Higgs mass.  
To derive Higgs mass bounds we first focus on a stable UV-potential
with a global minimum at vanishing field values.  In the absence of
higher-dimensional couplings, the lower Higgs mass bound can be
determined by requiring the quartic coupling to vanish at the cutoff
scale $\Lambda$. Higher-dimensional couplings at the scale $\Lambda$
can take values of $\mathcal{O}(1)$ in units of the cutoff
scale. Indeed, this is the generic situation that one would expect
when examining an effective field theory such as the Standard Model
close to its cutoff scale.   
In the
left panel of Fig.~\ref{fig:deltamhplot} we first confirm that the
observed Higgs mass around $125~\gev$ corresponds to the ultraviolet
boundary condition $\lambda_4=0$ at $\Lambda \approx
10^{10}~\gev$. Shifting this boundary condition closer to the Planck
scale and hence allowing for a fully stable potential would require a
Higgs mass close to $130~\gev$ with a fixed top
mass~\cite{shaposhnikov2012}.  We also show several choices of
ultraviolet boundary conditions. Those which predict smaller physical
Higgs masses than the choice $\lambda_i(\Lambda)=0$ allow us to
increase the ultraviolet cutoff.  According to Eq.\eqref{eq:stability}
a viable choice for a UV-stable potential including higher-dimensional
operators is for example $\lambda_4=-0.05$ and $\lambda_6=0.8$. The
corresponding Higgs mass stays below the conventional lower bound for
all values of $\Lambda$ and indeed gives a stable potential at the
Planck scale, with a dip into a pseudo-stable regime at intermediate
scales.

As described in detail in Sec.~\ref{sec:model} our calculation of the
Higgs mass relies on a numerically convincing, but nevertheless a toy
model. Special care is required when translating the computed shift of
the Higgs mass at fixed cutoff to the Standard Model.  We nevertheless
conclude that shifts at the level of $1 - 5 \%$ seem viable. To see
this we study the relative shift in the Higgs mass between the
perturbative Standard Model and including the dimension-6
self-coupling in Fig.~\ref{fig:deltamhplot}. It is given by
\begin{equation}
\Delta_{\mh}(\Lambda) = 
\frac{\mh^{(\lambda_i(\Lambda)\neq0)} - \mh^{(\lambda_i(\Lambda)=0)}}{\mh^{(\lambda_i(\Lambda)=0)}} \; .
\label{eq:def_deltamh}
\end{equation}
The dependence on $\Lambda$ is induced by the choice of ultraviolet
cutoff scale at which we define the boundary conditions for
$\lambda_4$ or $\lambda_6$. Negative values imply that the Higgs mass
resulting from finite ultraviolet values of $\lambda_4$ and
$\lambda_6$ lies below the conventional lower bound without exhibiting
instabilities in the potential in the UV or at the weak scale. We
emphasize, however, that large negative $\Delta_{\mh}$ go along with
pseudo-stable scenarios that deserve more detailed investigations.

For all choices of UV-couplings we observe that $|\Delta_{\mh}|$
decreases with increasing $\Lambda$. This can be understood from the
following argument~\cite{christof_fp}: including higher-dimensional
couplings can allow us to extend the renormalization group flow toward
the ultraviolet by several orders of magnitude, \ie the cutoff scale
is increased beyond the naive estimate determined by $\lambda_4(\Lambda) =0$. This
corresponds to shifting the curve $\mh(\Lambda)$ in the left panel of
Fig.~\ref{fig:deltamhplot} to the left by some amount $\Delta
\Lambda$.  The cutoff dependence of $\mh(\Lambda)$ flattens toward
higher cutoff scales, thus such a shift is less effective for large cutoffs, and correspondingly $\Delta_{\mh}$ decreases in this
regime.

While the achieved shifts in the allowed Higgs masses seem rather
small, we emphasize that with present uncertainties in the
experimental input parameters absolute stability of the Higgs
potential without $\lambda_{6}$ terms is disfavored at $\sim
3\sigma$~\cite{giudice}. Our shift of the Higgs mass limit should be viewed
as a shift in the central value which significantly reduces the
tension between the measured data and the possibility that the Higgs
potential is stable up to the Planck scale.\bigskip

So far, we have restricted ourselves to absolutely stable bare
potentials with a minimum at vanishing field values.  Next, we turn to
bare potentials bounded from below, but showing two minima.
Presumably, this property also holds for the effective potential.
However, since we employ a polynomial expansion of the potential we
can only study the renormalization group flow around the electroweak
minimum reliably. Future studies based on a different expansion will
have to shed further light on the global renormalization group flow of
the effective potential.

As already discussed in Sect.~\ref{sec:stability}, even a very small
positive value of $\lambda_{6}$ is sufficient to ensure stability.
However, the point $H=0$ is typically only meta-stable in this case.
The phenomenological viability of this scenario
requires that tunneling between the minima is sufficiently slow.  To
ensure this we enforce the conservative choice (cf. App.~\ref{app:tunnel}),
\begin{equation}
\lambda_{4}(k)>-0.052,\qquad \forall k.
\end{equation}
With this choice, the limit therefore essentially reduces to the usual
longevity limit for the meta-stable region as obtained in the absence
of any $\lambda_{6}$. 

Our numerical study based on a more advanced approximation as described in App.~\ref{app:compare} shows that already with moderately small values $\lambda_6\approx 0.25$ the cutoff scale can be increased by approximately two
orders of magnitude while retaining the full stability of the
electroweak vacuum. This can be seen from the
green
region I in
Fig.~\ref{fig:lambda6cutoff}. Increasing the possible cutoff scales by further orders of magnitude is 
difficult since
there is a strongly infrared-attractive pseudo-fixed point at
$|\lambda_6|\approx 0$ for those scales, see
Sec.~\ref{sec:fixed_point} below. Next, there is a large pseudo-stable
region 
(blue)
where the UV-potential as well as the low-energy
effective potential are stable, but our polynomially
expanded potentials exhibit further minima at intermediate scales.  As
already explained, this is beyond the strict validity of our
approximation and further studies are needed. Already relatively small
values of $\lambda_{6}$ are sufficient to stabilize the UV-potential
--- although not the minimum at $H=0$ --- up to the Planck scale,
since $\lambda_{4}$ is negative, but its absolute value remains quite
small.  Finally, in the 
red
region~III the UV-potential is already
meta-stable, as might be the effective potential. In this region the
cutoff can easily take values beyond the Planck scale even for tiny
values of $\lambda_{6}$, because the tunneling rates are always small
enough to guarantee the longevity of the electroweak vacuum.

If we are well into region I, our RG flow indicates that the effective potential at $k\approx 0$ is stable. Our approximation should be reliable in this region. The fate of the pseudo-stable region II is not quite as clear. The appearance of the second minimum at intermediate values of $k$
hints at the possibility that the effective potential at $k\approx 0$ might be meta-stable. However, to fully establish this feature requires
an approximation that goes beyond the local polynomial expansion we have used.
Finally, in region III it is likely that the effective $k\approx 0$ potential will be meta-stable.

\subsection{Fixed-point structure}
\label{sec:fixed_point}

\begin{figure}[t]
\includegraphics[width=\linewidth]{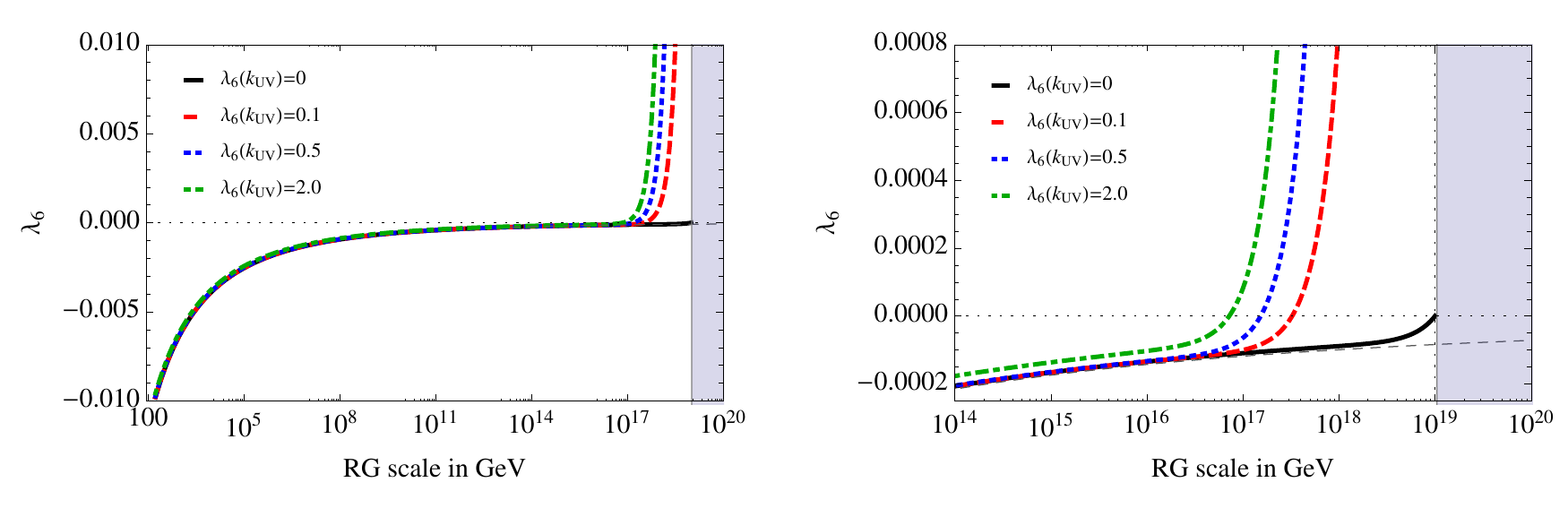}
\caption{\label{fig:toploop} Running $\lambda_6(k)$ along RG trajectories
  with $\lambda_6(k_\text{UV})=0$ (black solid), $\lambda_6(k_\text{UV})=0.1$
  (red dashed), $\lambda_6(k_\text{UV})=0.5$ (blue dotted),
  $\lambda_6(k_\text{UV})=2$ (green dot-dashed) and the pseudo-fixed point
  (thin grey dashed).}
\end{figure}

To understand the behavior of higher-order couplings below the cutoff
scale $\Lambda$, it is useful to analyze the beta functions in a
little more detail. For couplings $\lambda_6$ and higher, they have
the form
\begin{equation}
\beta_{\lambda_n} = (n-4) \lambda_n + \eta \lambda_n + c \; .
\end{equation}
The first term reflects the canonical dimension of the operator.  The
second term is a generalized anomalous dimension term, i.e., $\eta$ is independent of $\lambda_n$. We also include
contributions which are strictly speaking not an anomalous dimension,
but are nevertheless linear in the coupling. For example,
$\beta_{\lambda_6}$ contains such a term scaling like $\lambda_4
\lambda_6$. Finally, $c$ is a contribution from other couplings and
hence independent of $\lambda_n$. For instance, top quark fluctuations
contribute proportional to $y_t^n$.  The beta function for $\lambda_6$
thus features terms at most linear in $\lambda_6$. It is 
straightforward to find a pseudo-fixed point
\begin{alignat}{5}
\lambda_{6\, \ast}= - \frac{c}{2+\eta}, 
\qquad \text{with} \quad
c&= \frac{2 n_y N_c y^6 - 108\lambda_4^3}{16\pi^2}, 
\notag \\
\eta&=\frac{6 n_y N_c y^2 + 90\lambda_4}{16\pi^2} \; ,
\label{eq:fixed_point}
\end{alignat}
where the values for $c$ and $\eta$ can be read off from
Eq.\eqref{eq:betas_toy}. The pseudo-fixed point shows a remarkably small scale dependence, since the running of $\lambda_4$ and $y$ is also comparatively small.  In fact, the ratio $\lambda_4/y^2$ also approaches
a fixed point~\cite{christof_fp}.  We neglect a possible
contribution from $\lambda_8$.  Because the anomalous dimension $\eta$
is small, the canonical dimension renders the fixed point strongly
infrared attractive. In Fig.~\ref{fig:toploop} we show that arbitrary
starting values for $\lambda_6$ quickly converge toward the
pseudo-fixed point. At low scales, $\lambda_6$ follows the rapidly
growing top Yukawa coupling.  As expected from the sign of the
fermionic contribution in Eq.\eqref{eq:fixed_point}, $\lambda_6$
becomes negative. This does not signal an instability, as
higher-dimensional couplings also follow the value dictated by the
Yukawa coupling and the fermion determinant contributes positively to
the effective potential. The stability of the effective potential
including top quark effects is thus guaranteed by our finite
ultraviolet cutoff.\bigskip

In the perturbative approach, the explicit $\lambda_6$ dependence is
not taken into account. However, some of the corresponding effects are
encoded in higher loop diagrams~\cite{Codello:2013bra}. This can be
accounted for by setting $\lambda_6$ to its running pseudo-fixed point
value~\cite{christof_fp}. In Fig.~\ref{fig:toploop} we also show the
running fixed-point value from Eq.\eqref{eq:betas_toy}, \ie the
effective value of $\lambda_6$ included in loop effects in our
definition of $\beta_{\lambda_4}$. To understand why the $\lambda_n,\,
n>4$ follow their respective pseudo-fixed point values on all scales
in the perturbative approach, one should keep in mind that the cutoff
is sent to infinity in this case.  In an asymptotically free theory
$\lambda_6$ would vanish in the far ultraviolet and correspond to a
UV-repulsive direction.  Therefore, to stay on an asymptotically free
renormalization group trajectory would require a fine-tuned value
of $\lambda_6$ in the infrared, such that one would lie exactly on the
UV-critical surface. This corresponds to setting $\lambda_6$
essentially equal to the running fixed point value.
Because the Standard Model is not asymptotically free, there is no
reason to require that $\lambda_6$ should be set to its running
pseudo-fixed point value in the ultraviolet. Toward the infrared, it
will still approach the same value. In an effective theory $\lambda_6$
can be chosen independently of the pseudo-fixed point at $\Lambda$. A
given ultraviolet completion will actually dictate the ultraviolet
boundary condition.  A similar statement applies to all
higher-dimensional couplings.

\section{Models for high-scale physics}
\label{sec:light}

Starting with an effective field theory at a finite scale $\Lambda$,
the corresponding action can depend on many additional free parameters
in the top--Higgs sector.  An example is given by $\lambda_6$ as
studied in Sec.~\ref{sec:gaugemodel}. These parameters can have a sizeable
influence on the stability of the Higgs potential at high energy
scales. In Sec.~\ref{sec:higgsmass} we demonstrated that the interplay
of just the $\lambda_6$ term with the Higgs quartic coupling shifts
the limits on the measured Higgs mass from naive stability arguments
by several per-cent.  Let us now address the key question as to
whether such an effect can be generated in a particle physics model,
by integrating out heavy states. Depending on the choice of cutoff
scale, we would of course expect that the microscopic model could also
contain gravitational degrees of freedom. Here, we will focus on a
much simpler toy model to demonstrate the presence of higher-order
couplings at $\Lambda$.

\subsection{Heavy scalars}
\label{sec:heavy_scalars}

As a first step, let us check
if additional heavy scalars can provide
a model for new physics above $\Lambda$, which
features
stable potentials with $\lambda_6(\Lambda)>0$ and $\lambda_4(\Lambda)<0$.  In
the simplest scenario, the cutoff scale $\Lambda$ corresponds to the
mass scale of additional states, which are coupled to the Standard Model.
For scales $k>\Lambda$ they contribute to the renormalization group flow
of the SM couplings shown in Eq.\eqref{eq:betas_toy}.  To compute the
allowed Higgs masses in such a scenario, we do not need to consider a
 UV-complete theory for the heavy states beyond $\Lambda$.
Instead, the model can come with an inherent cutoff,
$\Lambda_\text{BSM} \gg \Lambda$. This corresponds to a hierarchy of
effective theories, in which the Standard Model is superseded by a
model containing heavy scalars, which again will be embedded in a more
fundamental model close to the Planck scale.\bigskip

In our simple model we couple the additional scalar to the Standard
Model through a Higgs portal~\cite{portal} added to the effective potential of
Eq.\eqref{eq:potential_toy2},
\begin{equation} 
\Delta V_\text{eff} (k)
= \lambda_{HS}(k)\frac{H^2}{2} \frac{S^2}{2} 
 + m_S(k)^2 \frac{S^2}{2} 
 + \lambda_S(k) \frac{S^4}{4} \; .  
\label{eq:lag_portal}
\end{equation}
Due to the reflection symmetry $S \rightarrow -S$, the heavy scalar is
stable and could constitute (a part of) the dark matter relic density. 
We assume that this $\mathbb{Z}_2$ symmetry remains unbroken.
The additional massive scalar field adds new loop terms to the beta
functions for $\lambda_4$ and $\lambda_6$, contributing only for
$k>m_S$. To decouple the massive modes below $m_S$ we include
threshold terms of the form $1/(1+ m_S^2/k^2)^n$ with an appropriate
power $n$. Including these threshold effects, the loop terms can be
calculated in the FRG scheme, as shown in App.~\ref{app:compare}. If
we allow for $N_S$ mass-degenerate scalar fields with the same Higgs
portal interaction, the one-loop beta functions of our toy model,
Eq.\eqref{eq:betas_toy}, become
\begin{alignat}{6}
\beta_{\lambda_4} &= \frac{1}{8\pi^2} \;
 \left[ - n_y N_c y^4+2 n_y N_c y^2\lambda_4 
        + 9\lambda_4^2 
        - \frac{15}{4} \lambda_6 
        + c_\lambda g_F^4
        + \frac{N_S \lambda_{HS}^2}{4(1+m_S^2/k^2)^3} 
 \right] 
\notag \\
\beta_{\lambda_6} &= 2\lambda_6 +\frac{1}{16\pi^2}
 \left[ 2 n_y N_c y^6
      + 6 n_y N_c y^2\lambda_6 
      - 108\lambda_4^3
      + 90\lambda_4 \lambda_6 
      - \frac{N_S \lambda_{HS}^3}{2(1+m_S^2/k^2)^4} 
 \right] \; .
\end{alignat}
The Higgs portal contributions to the beta functions follow the usual
pattern for bosonic fluctuations: $\beta_{\lambda_4}$ receives a
positive contribution, while for $\beta_{\lambda_6}$ the sign is
reversed, etc.  Most relevant to our argument is the negative sign in
$\beta_{\lambda_6}$, because it implies a growing coupling toward the
infrared, and thus a stabilizing effect. Looking at the effective
potential we can generalize the stabilizing effect of heavy scalars
beyond the leading terms in a polynomial expansion, namely
\begin{equation}
\frac{d}{d \log k} \frac{V_\text{eff}(k)}{k^4} 
\sim  \frac{1}{1+\lambda_{HS} H^2} \; .
\end{equation}
Through the overall positive sign heavy scalars indeed decrease the
value of the effective potential toward the infrared. Furthermore, the
strength of the effect depends on the value of the Higgs field, and is
largest at small field values of $H$. In the following, we will
restrict ourselves to $\lambda_4$ and $\lambda_6$, but keep in mind
that higher-dimensional couplings will be generated with alternating
signs. While truncating this series at finite order could suggest
either stability or instability, the contribution to the effective
potential generated by heavy scalars remains stable, and features a
minimum at vanishing field.\bigskip

\begin{figure}[t]
\includegraphics[width=1.0\textwidth]{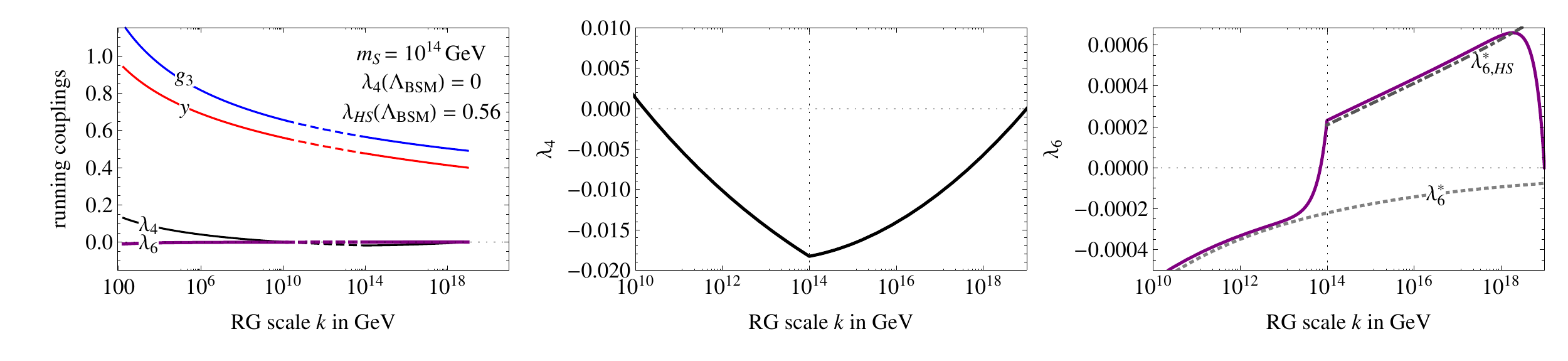}
\includegraphics[width=1.0\textwidth]{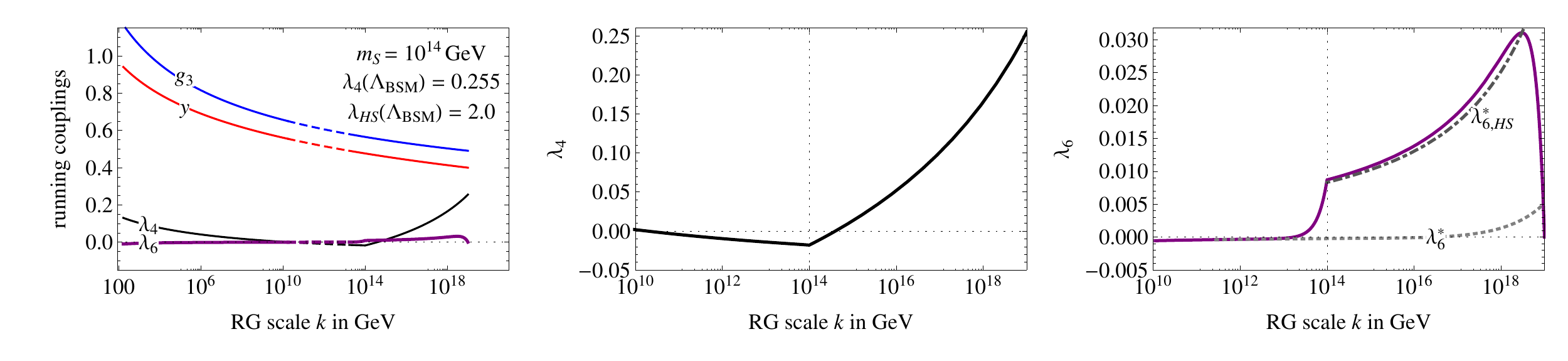}
\includegraphics[width=1.0\textwidth]{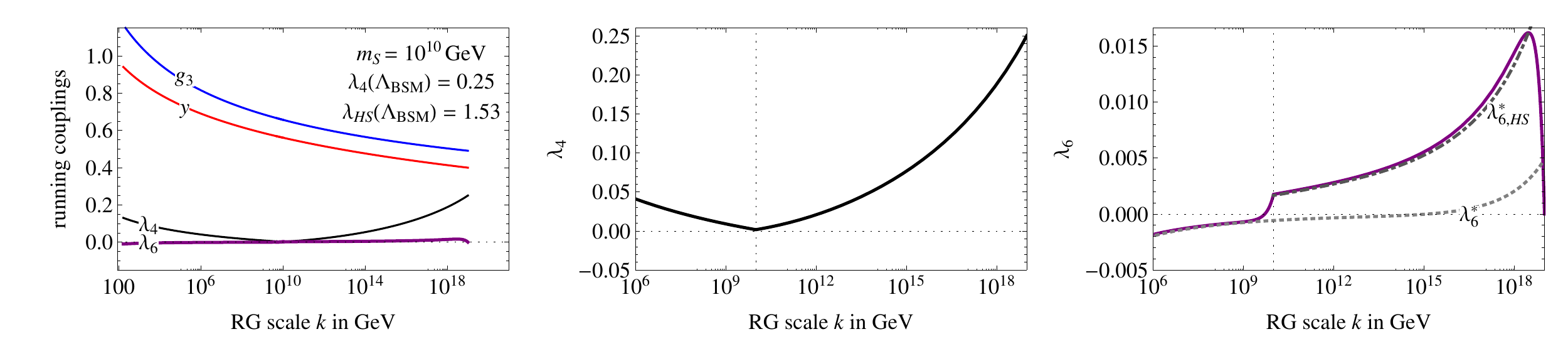}
\caption{\label{fig:GHYdm} Running couplings in the presence of a
  Higgs portal coupling $\lambda_{HS}$ and $N_S=3$. In the top row the boundary
  condition is $\lambda_4(\Lambda_\text{BSM})=0$, while in the second
  row it is replaced by a finite value.  In the bottom row we
  illustrate a moderately big $\lambda_{HS} \sim 1$ at the Planck
  scale, which is sufficient for an absolutely stable potential at all
  scales. In the right panels, we show the pseudo fixed point value for $\lambda_6$ in the presence of heavy scalars, $\lambda_{6, HS}^{\ast}$, and without heavy scalars, $\lambda_6^{\ast}$.}
\end{figure}

For approximately constant $\lambda_{HS}(k)$ the value of $\lambda_4(\Lambda)$
will depend on $-\log (\Lambda_\text{BSM}/ \Lambda)$. Thus, the bare value
$\lambda_4(\Lambda_\text{BSM})$ will determine $\lambda_4(\Lambda)$.
On the other hand, higher-dimensional couplings such as $\lambda_6$
reach values which are independent of $\Lambda_\text{BSM}$, as
expected from their canonical dimensionality:
As in the case of the pure Standard Model, the infrared value of
$\lambda_6$ in the presence of a heavy scalar is determined by a
pseudo-fixed point.  In Fig.~\ref{fig:GHYdm} we demonstrate that the
behavior of $\lambda_6$ is completely determined in terms of this
strongly IR-attractive pseudo-fixed point, making the value of
$\lambda_6(\Lambda_\text{BSM})$ irrelevant for weak-scale observables, as long as $\Lambda \ll \Lambda_{\rm BSM}$.
All three scenarios shown yield the correct values $\mh = 125~\gev$
and $m_t^\text{(pole)} = 173~\gev$. The value of $\lambda_6(\Lambda)$
only depends on the relevant and marginal couplings at that scale. As
expected from universality arguments, $\lambda_6$ forgets the dynamics
between $\Lambda_{\text{BSM}}$ and $\Lambda$, different from the
marginally relevant coupling $\lambda_4$, which does depend on the
details of the dynamics at $\Lambda_\text{BSM}$.

Comparing the full solution to the pseudo-fixed point trajectory in
and beyond the Standard Model we indeed observe that setting
$\lambda_6(\Lambda) = \lambda_{6\, \text{SM}}^{\ast}$ (and
correspondingly for all higher-dimensional operators) is too
restrictive. Instead, the value of $\lambda_6$ is determined by an
interplay of two pseudo-fixed points: As long as $k > \Lambda$,
$\lambda_6$ is determined by the pseudo-fixed point in the presence of
new physics. Below the scale of new physics, $\lambda_6$ undergoes a
transition to the pseudo-fixed point determined by the relevant and
marginal Standard Model couplings. A caveat is that this analysis
applies only to situations with small anomalous dimensions, \ie in a
perturbative regime.\bigskip

We can now determine the conditions under which $\lambda_4(\Lambda)$
and $\lambda_6(\Lambda)$ assume values that according to
Sec.~\ref{sec:higgsmass} yield Higgs masses below the conventional
lower bound.  First, we re-iterate our earlier observation that
seeming instabilities in the potential in
Fig.~\ref{fig:GHYdm} are artifacts of our truncation to $\lambda_4$
and $\lambda_6$. To fix $\lambda_4$ and $\lambda_6$ we set
$\lambda_4(\Lambda_\text{BSM})$ such that the Higgs mass comes out
correctly.  We then solve the pseudo-fixed point equation for
$\lambda_6$, which determines its value at $\Lambda$.  We obtain
\begin{equation}
\lambda_6(\Lambda) 
= \frac{-12 y^6(\Lambda) +216 \lambda_4^3(\Lambda) + N_S \lambda_{HS}^3(\Lambda)}
       {4 \left(16 \pi^2 + 9 y^2(\Lambda)+45 \lambda_4(\Lambda) \right)} \; .  
\label{eq:lambda6_scalars}
\end{equation}
A positive top Yukawa and a negative Higgs quartic coupling, which is
the case of interest for the Higgs mass bounds, each reduce the
pseudo-fixed point value. As the top Yukawa coupling grows toward the
infrared, this negative contribution increases. Accordingly, a
sufficiently large and positive value of $\lambda_{HS}$ is needed for
a positive fixed-point value $\lambda_6^*$, which is then depleted
toward lower scales. In contrast, $\lambda_4$ is driven to
increasingly negative values by the fluctuations of the heavy scalar.
This implies that while the generated value of $\lambda_6$ provides a
potential that is bounded from below, it is not always large enough to
yield a potential with a minimum at vanishing field values. The size
of the quadratic Higgs term $\mu^2$ still decides whether the
potential at the cutoff scale $\Lambda$ is meta-stable, or features a
global minimum at vanishing Higgs field.

It is nevertheless possible to generate initial conditions
corresponding to a stable bare potential with a minimum at a vanishing
field value.
 As an example, we use $\Lambda = 10^{14}~\gev$, where
$y(\Lambda) =0.476$ and $\lambda_4(\Lambda) =-0.017$ give a physical
Higgs mass of $125~\gev$. This value of $\lambda_4(\Lambda)$ can be
reached by adjusting $\lambda_4(\Lambda_\text{BSM})$.  To obtain a
UV-stable potential with the measured Higgs mass value, we need
\begin{equation} 
N_S \lambda_{HS}^3 \gtrsim 24 \; , 
\end{equation}
for example corresponding to $\lambda_{HS} \sim 2$ and 3 additional
scalars.  This large value of $N_S \lambda_{HS}^3$ is needed to
compensate for the factor 216 in front of the $\lambda_4$ term in
Eq.\eqref{eq:lambda6_scalars}, which arises from the combinatorics of
the respective scalar diagrams. Accordingly, a new physics scenario in
which the new states are not combinatorically disfavored in comparison
to the Higgs can accommodate larger values of $\lambda_6$ without
needing large numbers of new states and/or large couplings.

In our example, the large value of the Higgs portal coupling implies
that the heavy scalar sector will be driven toward a Landau pole not
far above $\Lambda_\text{BSM}$.
In such simple models one could therefore conclude that a UV-stable
potential for the Higgs sector can be generated at the cost of a
non-perturbative regime not far above $\Lambda_\text{BSM}$. 

\subsection{Heavy fermions}
\label{sec:heavy_fermions}

An obvious open question is the effect of heavy fermions. If their
mass is generated through symmetry breaking at the electroweak scale,
they will have a large Yukawa coupling. For example models featuring a
heavy chiral fourth generation show a significantly reduced value of
the possible cutoff scale $\Lambda$ since the extra fermions would
just add to the problematic effect of the top quark.  Furthermore,
such models are experimentally excluded through the Higgs coupling
measurements at the LHC.

We consider a model where the additional fermions are heavy, but their
Yukawa coupling is small. In this setting we rely on an unspecified
symmetry breaking mechanism at a high scale, affecting only the
additional fermions. In our simple model we include such a mass term
without discussing its possible origin. While this is straightforward
in the present $\mathbb{Z}_2$ model, in spite of the mass term
explicitly breaking the $\mathbb{Z}_2$ symmetry, an embedding of such a
mechanism in the Standard Model with its $SU(2)_L$ gauge invariance is
less clear.  The relevant Lagrangian terms for $N_\eta$ heavy fermions
then take the form
\begin{equation}
\mathcal{L}_\eta \supset y_\eta H \bar{\eta}\eta + m_\eta^2 \bar{\eta} \eta \; . 
\end{equation}
Alternatively, we could consider a model where the additional
fermions are singlets under the Standard Model gauge groups. In that
case they can have a mass even in the unbroken phase. Their Higgs
couplings would correspond to a dimension-5 operator of the form $H^2
\bar{\eta} \eta$.\bigskip

Let us again analyze the induced potential at $\Lambda = m_\eta$ in
terms of the pseudo-fixed point for $\lambda_6$, which instead of
Eq.\eqref{eq:lambda6_scalars} now gets the additional contribution
\begin{equation}
\lambda_6(\Lambda) = \frac{-12 y^6(\Lambda)+216 \lambda_4^3(\Lambda)+ N_S \lambda_{HS}^3(\Lambda) -12 N_\eta y_\eta^6(\Lambda)}{4 \left(16 \pi^2 + 9 y^2(\Lambda)+45 \lambda_4(\Lambda) +9N_\eta y_\eta^2(\Lambda)\right)}.
\end{equation}
The fermion $\eta$ induces a negative contribution to
$\lambda_6$. In general, its contribution to $\lambda_n$ will be
positive for $n/2$ even, and negative for $n/2$ odd. This means that
heavy fermions will generically make it hard to reach sizeable
positive values of $\lambda_6$, but they might be of interest for
cases where, e.g., $\lambda_8>0$ stabilizes the potential.\bigskip

Alternatively, heavy fermions can be important 
in a scale invariant theory in the far ultraviolet,
$V(\Lambda_\text{BSM})=0$. Here, fermions help to control the induced
value of $\lambda_4(\Lambda)$. In a simple analysis the masses of
heavy bosons and fermions split the new physics scale $\Lambda$ into
two thresholds $m_S= \Lambda_S$ and $m_\eta = \Lambda_\eta$, where we assume $m_S < m_{\eta}$. The
induced value of $\lambda_4$ for $\Lambda_S< \Lambda_{\eta}$ now reads
\begin{equation}
\lambda_4(\Lambda_S) 
=  \lambda_4(\Lambda_\text{BSM}) 
 +\left( \frac{3}{8\pi^2}y^4(\Lambda_S) 
        -\frac{N_S}{32 \pi^2}\lambda_{HS}(\Lambda_S)^2 
   \right) \log \frac{\Lambda_\text{BSM}}{\Lambda_S}
  +\frac{N_\eta}{8\pi^2}y_\eta^4(\Lambda_{\eta}) 
    \log \frac{\Lambda_\text{BSM}}{\Lambda_\eta} \; ,
\end{equation} 
where we assume a slow running in the Yukawa and the portal
couplings to the heavy scalars. A hierarchy $m_\eta> m_S$ implies
that the value of the induced coupling $\lambda_6$ will be independent
of $y_\eta$ and instead be determined by the pseudo-fixed point at
$\Lambda_S$. In this framework we can construct models where $\lambda_4$
is small and negative and $\lambda_6 = \mathcal{O}(1)$, without the
need for large fermion and boson numbers or large couplings. 

\subsection{Non-perturbative new physics}
\label{sec:heavy_general}

\begin{figure}[t]
\begin{center}
\includegraphics[width=0.40\linewidth]{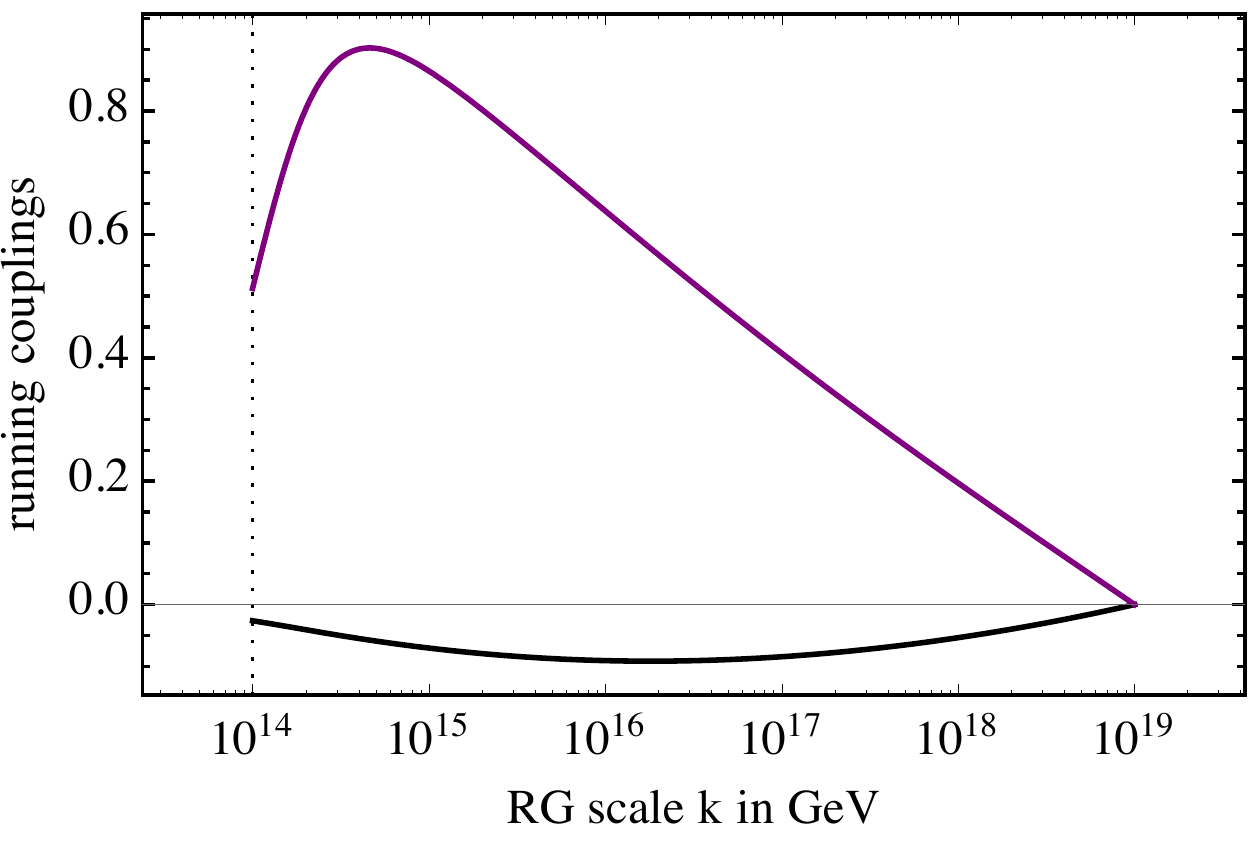} 
\end{center}
\caption{We show $\lambda_4$ (black) and $\lambda_6$ (purple),
  obtained using the model beta functions from Eq.\eqref{eq:betasNP},
  which include new-physics effects beyond $\Lambda = 10^{14}~\gev$
  (dotted vertical line), induce these couplings and alter their
  canonical scaling dimensionality.}
\label{fig:lambdasNP} 
\end{figure}

In the last two sections we have shown that simple models with
additional heavy fermions and bosons allow us to generate initial
conditions for the renormalization group flow which yield
low Higgs masses based on a
UV-stable potential.  The main challenge is to avoid the non-perturbative
regime beyond $\Lambda_{\rm BSM}$ while generating coupling values $|\lambda_6|> |\lambda_4|$.
In general, the effect of new physics on the beta functions of the
running couplings in the Higgs sector follows two different patterns:
\begin{enumerate}
\item terms independent of the Higgs self-couplings induce these
  couplings, even if they vanish at some scale $\Lambda_\text{BSM}$;
\item terms which change the scaling of the Higgs self-couplings by
  inducing an anomalous dimension $\eta_\text{NP}$.
\end{enumerate}
This means we can write the beta functions for $\lambda_4$
and $\lambda_6$ in the presence of new heavy states as
\begin{alignat}{5}
\beta_{\lambda_4} &= \frac{1}{8\pi^2} \;
 \left[ - n_y N_c y^4+2 n_y N_c y^2\lambda_4 
        + 9\lambda_4^2 
        - \frac{15}{4} \lambda_6 
        + c_\lambda g_F^4 + c_{4\, \text{NP}} \right] + \eta_{4\, \text{NP}} \lambda_4
 \notag \\
\beta_{\lambda_6} &= 2\lambda_6 +\frac{1}{16\pi^2}
 \left[ 2 n_y N_c y^6
      + 6 n_y N_c y^2\lambda_6 
      - 108\lambda_4^3
      + 90\lambda_4 \lambda_6 + c_{6\, \text{NP}}  \right]+ \eta_{6\, \text{NP}} \lambda_6,
\label{eq:betasNP}
\end{alignat}
where $c_{4/6\, \text{NP}}$ and $\eta_{4/6\, \text{NP}}$ encode the
effects of the microscopic degrees of freedom beyond the Standard
Model.  Again, it is useful to analyze the running of $\lambda_6$ in
terms of a pseudo-fixed point of the kind shown in
Eq.\eqref{eq:fixed_point},
\begin{alignat}{5}
\lambda_{6\, \ast}= - \frac{c}{2+\eta} 
\qquad \text{with} \quad
c&= \frac{2 n_y N_c y^6 - 108\lambda_4^3 + c_{6\, \text{NP}}}{16\pi^2}
\notag \\
\eta&= \frac{6 n_y N_c y^2+ 90\lambda_4}{16\pi^2} +\eta_{6\, \text{NP}} \; .
\label{eq:fixed_point2}
\end{alignat}
For example, we can start with vanishing values for all couplings at
$\Lambda_\text{BSM}$. We can induce a negative value of
$\lambda_4$ with increasing magnitude and a positive value of
$\lambda_6$, if we choose $c_{4\, \text{NP}}>0$ and $c_{6\,
  \text{NP}}<0$ and $|c_{6\, \text{NP}}|>2 n_y N_c y^6$.  The
pseudo-fixed point moves toward smaller values as $\lambda_4$ grows
in magnitude toward the infrared.  As long as $\eta_{6\, \text{NP}}$
is small the pseudo-fixed point stays strongly infrared
attractive.\bigskip

Let us now broaden our approach and allow for new physics in a
non-perturbative regime, such that large anomalous dimensions of
either sign can exist. An example could be asymptotic
safety~\cite{asymptotic_safety}. We choose a scenario
where $\eta_{6\, \text{NP}} \simeq -2$, \ie, the scaling dimension of
the higher-dimensional coupling changes from irrelevant to marginal, or
even relevant. In such a case, the pseudo-fixed point will not be
strongly infrared attractive anymore. Accordingly, new physics can
induce a sizeable value of $\lambda_6$, that is not depleted between
$\Lambda_\text{BSM}$ and $\Lambda$. From Eq.\eqref{eq:fixed_point2}
it is clear that cancelling the canonical dimensionality by a large
anomalous dimension can lead to significantly enhanced values for
$\lambda_6^*$.

As an example, we set $c_{4\, \text{NP}}=9/4$, $c_{6\, \text{NP}} =
-27/2$, $\eta_{4\, \text{NP}}= 0$, and $\eta_{6 \, \text{NP}}=-2$ and
assume that the effect of new physics appears beyond $\Lambda =
10^{14}~\gev$. We then obtain $\lambda_4(\Lambda) = -0.026$ and
$\lambda_6(\Lambda) = 0.51$, as shown in
Fig.~\ref{fig:lambdasNP}. This does provide a UV-stable potential with
Higgs masses significantly below the conventional lower bound.

\section{Conclusions}
\label{sec:conclusions}

With the discovery of the Higgs boson the Standard Model describing
the interactions between fundamental states in terms of a gauge theory
is complete.  The crucial question becomes whether the Standard Model
itself gives us a hint where new physics must appear~\cite{1000Higgs}.
One possible clue is the behavior of the Higgs potential at high
energy scales and large field values. A perturbative analysis of the
Higgs potential in the Standard Model indicates that it loses its
stability around $10^{10}~\gev$, well below the Planck scale. We have
investigated the robustness of this claim in the presence of
higher-dimensional operators in the effective Higgs potential, as one
would expect in theories with a physical UV-cutoff.  Because of their
canonical dimension the effects of these higher-dimensional operators
will vanish quickly toward the infrared, but nevertheless can play a crucial role in the vicinity of the cutoff.  We identify several
scenarios:
\begin{itemize}
\item{} The UV-potential including higher-dimensional couplings with
  $\lambda_4<0$ is stable, as is the effective potential.  The
  electroweak minimum remains the global minimum throughout the RG
  evolution. In this case, already the $H^6$ operator allows us to
  extend the UV-cutoff by about two orders of magnitude. Further
  higher-dimensional operators may increase the cutoff scale further. 
\item{} The UV-potential and the effective potential are stable with
  their respective global electroweak minima. However, during the RG
  evolution this minimum seems only meta-stable for an intermediate range
  of scales.  For conclusive statements about this pseudo-stable
  scenario, a more complete treatment of the RG evolution of the
  effective potential is necessary. We expect that part of this
  parameter space belongs to the class of fully stable scenarios
  whereas other parts may feature a real meta-stability.  Taking the
  scenario literally, the barrier separating the two minima of the
  $k$-dependent potential is typically of the order $k$.
\item{} The UV-potential is meta-stable, featuring a global minimum at
  some high field value. This can be achieved for an arbitrarily small
  initial $\lambda_{6}$, and therefore no instability occurs even for
  arbitrarily large negative $\lambda_4$.  Implementing the constraint
  that the electroweak minimum has to be sufficiently long-lived, this
  essentially reduces to the standard discussion of a viably
  long-lived meta-stable region.
\end{itemize}
\bigskip

Higher-dimensional operators affect the constraints on the Higgs mass
between $\sim1~\gev$ for the stable potential to $\sim4~\gev$ for a
pseudo-stable setup.  In the presence of an ultraviolet cutoff, the
size of higher-dimensional couplings is an inherent ambiguity of the
Standard Model.  To include higher-dimensional operators in our setup
we have used non-perturbative functional renormalization group
techniques in the presence of a finite UV-cutoff.  Neglecting
higher-dimensional operators, our approach and the usual perturbative
approach agree in the domain where perturbation theory
with a massless regularization scheme is applicable, \ie, for energies
and field values much smaller than the UV-cutoff $\Lambda$. With
regard to higher-dimensional operators, perturbation theory assumes
very specific and small values, essentially those generated by the
Standard Model degrees of freedom and encoded in higher-order
effects. Our non-perturbative treatment allows for more general values
of the higher-dimensional couplings. 

The difference between the values of the couplings 
in
perturbation theory and our more general values is in principle
measurable. These effects become sizeable at energies where the new
dimensionless couplings are of order one. Restricting the higher-dimensional
couplings to be of order one at the cutoff scale (where they typically
have their largest values), limits the size of the achievable shifts, but even in the absence of
higher-dimensional operators, stability up to the Planck scale is only
disfavored by $\sim3\sigma$~\cite{giudice}. Small shifts as observed
in our analysis can significantly reduce this tension.\bigskip

Physics beyond the Standard Model predicts the size of the Higgs
self-interactions at the cutoff. We can ask what type of new physics
generates higher-dimensional operators of suitable size, such that
Higgs masses below the conventional lower bound can be obtained.  In general,
UV-completions featuring large anomalous dimensions can easily
generate sizeable higher-dimensional operators, stabilizing the Higgs
potential up to the Planck scale. Their effects can be linked directly
to the more general analysis in terms of higher-dimensional operators
contributing to the Higgs potential.

In weakly coupled models the answer is less obvious. Our approach to
new physics affecting vacuum stability is different from the usual
strategy of modifying the running of the Higgs potential through new
particles below $10^{10}~\gev$, as discussed in the Appendix. In the
main body of our paper we focus on very heavy new particles, coupled
to the Standard Model through renormalizable operators. They
modify the Higgs potential through induced higher-dimensional Higgs self-couplings.  Specifically, we have
investigated a simple extension of the Standard Model by heavy scalars
and fermions.  While it is possible to generate sizeable
higher-dimensional operators, the required parameter choices in our
simple model (couplings and number of fields) are on the border of
being non-perturbative. It remains an intriguing question, what kind
of new physics can generate higher-dimensional operators of a suitable
size, while remaining perturbative on all scales.

\acknowledgments 

We thank Christof Wetterich and Jan Martin Pawlowski for
discussions. TP would like to thank Christof Wetterich for giving him
the friendly--supportive Christof smile when he heard about this
project.  The work of AE is supported by an Imperial College Junior
Research Fellowship. This research was supported in part by Perimeter
Institute for Theoretical Physics. Research at Perimeter Institute is
supported by the Government of Canada through Industry Canada and by
the Province of Ontario through the Ministry of Research and
Innovation.  AE and RS thank the ITP at Heidelberg University, and MMS
the Perimeter Institute for hospitality during the course of this
work. MMS is supported by the grant ERC-AdG-290623.  HG and RS
acknowledge support by the DFG under grants GRK1523, Gi328/5-2
(Heisenberg program).

\clearpage
\appendix

\section{Light dark matter scalars}
\label{app:dark_matter}

\begin{figure}[!b]
\includegraphics[width=0.5\linewidth]{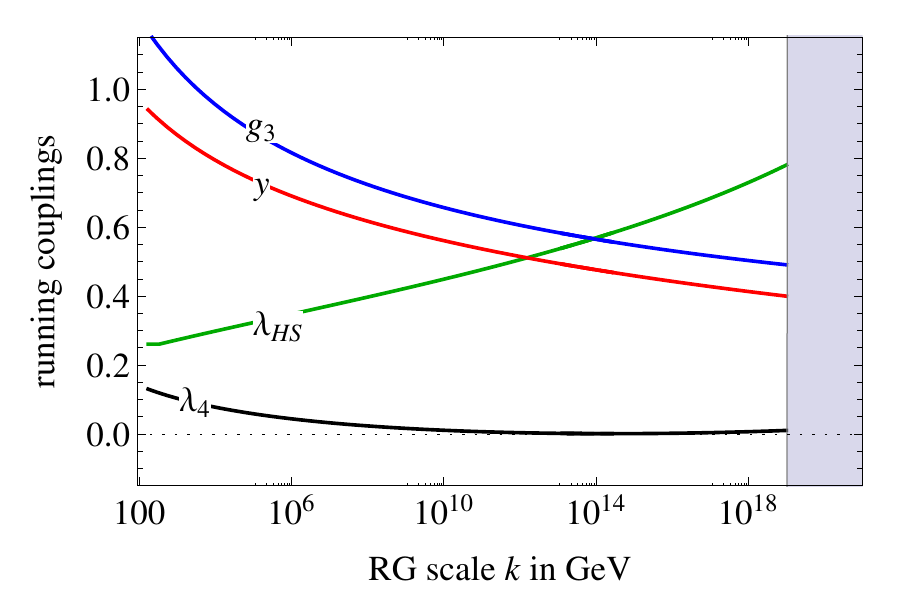}
\caption{\label{fig:runningSMplusdm}We show the running of the
  couplings employing Eq.\eqref{eq:betas_toy}, amended by an additional
  Higgs portal coupling $\lambda_{HS}$. }
\end{figure}

Different scenarios predicting new particles which modify the relation $m_H(\Lambda)$ from the
conventional lower bound can be distinguished by the ratio of their
mass to the cutoff $\Lambda$.  In the main body of this paper we focus
on heavy states with mass $m \gtrsim \Lambda$. Their fluctuations
induce an effective potential $V_\text{eff}(\Lambda)$ which
generically contains non-vanishing higher-dimensional terms.

A different scenario arises using light states with mass $m \gtrsim
v$. Their primary effect is to modify the running of (some of) the
Standard Model couplings between the electroweak scale and the cutoff
scale. Most importantly, they can alter the RG flow of the quartic
Higgs self-coupling, such that it remains positive all the way to the
cutoff $\Lambda$. One example is a gauge singlet, dark matter
scalar. It is coupled to the Standard Model through the Higgs portal
defined in Eq.\eqref{eq:lag_portal}, $\lambda_{HS} H^2 S^2/4$, but now
with a mass $m_S$ in the GeV-TeV
region~\cite{portal,eichhorn_scherer}. With a $\mathbb{Z}_2$ symmetry
protecting it from decaying and in the absence of a second VEV which
would lead to mixing with the Higgs scalar, the only additional
parameters of this model are $m_S$ and $\lambda_{HS}$.  This makes it
the arguably simplest model, which can accommodate the complete
observed dark matter relic density~\cite{Cline:2013gha}. From a field
theory point of view, the portal coupling cannot be large and negative
to avoid an unstable combined scalar potential. To yield the desired
effect on the Higgs mass bound it should be positive.\bigskip

The difference to the additional heavy scalar discussed in
Sec.\ref{sec:heavy_scalars} is that $\lambda_{HS}$ is now important at
scales $k < \Lambda$.  A sufficiently large $\lambda_{HS}$ can
altogether prevent $\lambda_4$ from becoming negative. In
Fig.~\ref{fig:runningSMplusdm} we update results from
Ref.~\cite{eichhorn_scherer} by including the effect of the strong
gauge coupling and approximating the effect of the weak couplings
through a fiducial contribution as defined in Eq.\eqref{eq:betas_toy}.
A moderately large $\lambda_{HS} \sim 0.78$ at the Planck scale is
indeed sufficient for an absolutely stable potential for the Higgs at
all scales. In our updated analysis we also guarantee the correct
weak-scale masses $m_h = 125~\gev$ and $m_t^\text{(pole)}= 173~\gev$.
The infrared value of $\lambda_{HS}$ in Fig.~\ref{fig:runningSMplusdm}
is compatible with a dark matter mass of $m_S = 340~\gev$, if
we require the new scalar to constitute the complete dark matter relic
density~\cite{Cline:2013gha}.

From this example we tentatively conclude that the positive
contribution of $\lambda_{HS}$ to $\beta_{\lambda_4}$ is sufficient to
reconcile the measured value of the Higgs mass with a stable Higgs
potential on all scales up to the Planck scale. As a crucial
phenomenological distinction to our investigation in the main part of
this paper, new physics states appear at low scales $m_S \gtrsim v$,
instead of at mass scales $m_S \sim \Lambda$.  Thus the new light
states can become experimentally accessible in dark matter searches of
at colliders in the near future. The particular region of parameter
space used in Fig.~\ref{fig:runningSMplusdm} will be probed by planned
upgrades of XENON~\cite{Aprile:2012zx} and LUX~\cite{Fiorucci:2013yw}.

\section{Tunnelling}
\label{app:tunnel}

In this appendix we briefly discuss estimates for the tunneling rates
in the presence of a $\sim \lambda_6\varphi^6$-term~\cite{branchina}.
The tunnelling rate can be calculated by determining the Euclidean
action of bubbles $S_\text{E}[\varphi_\text{bounce}]$, where
$\varphi_\text{bounce}$ is the bounce solution of the Euclidean
equations of motion for a real scalar field
\begin{equation}
\frac{d^2 \varphi_\text{bounce}}{dr^2}+\frac{3}{r}\frac{d\varphi_\text{bounce}}{dr} -V'(\varphi_\text{bounce})=0 \; ,
\end{equation}
with the boundary conditions
$\varphi_\text{bounce}(\infty)=\varphi_\text{bounce}=\varphi_{+}$ and
$\varphi_\text{bounce}'(0)=0$. The value $\varphi_{+}$ is the
location of the meta-stable minimum. The tunneling rate then becomes
\begin{equation}
\frac{\Gamma}{V}\sim e^{-S_\text{E}[\varphi_\text{bounce}]} \; .
\end{equation}

Let us now turn to the potential given by
Eq.\eqref{eq:potential_toy2}, namely
\begin{equation}
\label{tunnelingpotential}
V(\varphi)
=\frac{m^2_{\varphi}}{2} \, \varphi^2
+\frac{\lambda_4}{4} \, \varphi^4
+\frac{\lambda_6}{8 k^2} \, \varphi^6 \; .
\end{equation}
For the formation of a Bubble of size $\sim R$ one should evaluate the
scale dependent potential at the scale
$k\sim1/R$~\cite{Strumia:1998qq,Isidori:2001bm}. In the situation we
are interested in, the field values where the potential starts to be
lower than the electroweak vacuum near zero vev are quite large.
Accordingly we are interested in very small bubbles and $k\gtrsim
10^{10}~\gev$.  Neglecting higher-order terms, the potential would be
unstable with $m^2_{\varphi}>0$ and $\lambda_4<0$.  To stabilize the
potential we can add a $\lambda_6\varphi^6$ term with,
$\lambda_6>0$.\bigskip

We are mainly interested in an upper limit on the tunneling rate. It
therefore
makes  sense to approximate the potential
Eq.\eqref{tunnelingpotential} by a potential which is always lower
than the potential we actually have.  A simple choice is to simply
take the term
\begin{equation}
V_\text{approx}(\varphi)=\frac{1}{4}\lambda_4\varphi^4\leq V(\varphi).
\end{equation}
For such a potential the tunneling rate has been computed in
Ref.~\cite{Isidori:2001bm}.  In this case the bounce solution is,
\begin{alignat}{5}
\varphi_\text{bounce} &=\sqrt{\frac{2}{|\lambda_4|}}\frac{2R}{r^2+R^2} \notag \\
S_\text{E}[\varphi_\text{bounce}] &= \frac{8\pi^2}{3|\lambda_4|} \; .
\label{phi4bounce}
\end{alignat}
Because the approximate potential is scale invariant we actually have infinitely many bounces corresponding to different values of $R$, and in principle we have to take all of them into account.

To determine the pre-factor of the exponential one has to perform a
one-loop calculation around the background of the bounce
solution~\cite{Callan:1977pt}.  Doing this~\cite{Isidori:2001bm} one
finds for the tunnelling probability in a space time volume $V_4$
\begin{equation}
p\sim \frac{V_4}{R^4}\exp\left(-\frac{8\pi^2}{3|\lambda_4(1/R)|}\right).
\end{equation}
The scale dependence of $\lambda_4$ breaks the scale invariance and
differently sized bubbles have different actions.  To determine the
loop-correction one has to calculate the inverse of the determinant
over the spectrum of the fluctuation modes around the bounce
background. The zero-modes have to be treated with some extra
care. They correspond to integrations over the corresponding
collective coordinates. The volume factor arises from the translation
zero-modes and the corresponding integration over the space-time
volume.  The scale invariance of the approximate potential leads to an
additional zero-mode in the spectrum.  The integral over $dR/R$ is the
integration over the collective coordinate corresponding to this
zero-mode~\cite{'tHooft:1976fv}.  Taking all possible bubbles into
account we finally arrive at
\begin{equation}
\label{tunnelintegral}
p\sim \int \frac{dR}{R}\,\frac{V_4}{R^4}\exp\left(-\frac{8\pi^2}{3|\lambda_4(1/R)|}\right)
\end{equation}
For constant $\lambda_4(1/R)$ this integral is divergent for small
values of $R$.  Such small values of $R$ correspond to a large maximal
field values of the bounce solution, as can be seen from
Eq.\eqref{phi4bounce}.  Indeed we have,
\begin{equation}
\varphi_\text{bounce}(r=0)\sim \frac{1}{R} \; . 
\end{equation}
This tells us how the integral is regularized in presence of
$\lambda_6>0$ which stabilizes the potential. For field values
beyond the point where the potential has its minimum and starts to
rise again, the approximation of the potential by just the
$-\lambda_4\varphi^4/4$ term is not sensible anymore and we should
cut off the integral at this point at the latest.

In practice, the integral is typically dominated by a reasonably sized
region around the most negative values of $\lambda_4$ because of the
strong exponential suppression of the integrand for small
$|\lambda_4|$ in Eq.\eqref{tunnelintegral}. Even a small upturn in
$\lambda_4(1/R)$ typically  quickly overcomes any growth in the
$1/R^4$ factor.  Indeed a better approximation is probably to use an
effective quartic coupling
\begin{equation}
\lambda^\text{eff}_4=\lambda_4 + \frac{\lambda_6}{2 k^2} \varphi^2.
\end{equation}
Here, one can again see the stabilizing effect of a positive
$\lambda_6$. With increasing field value, \ie decreasing $R$ of the
bounce, $\lambda^\text{eff}_4$ becomes less negative effectively
cutting off the integral. More intuitively, an increasing $\lambda_6$
lowers the depth of the minimum, and thus lowers the tunneling
probability.\bigskip

Let us now determine the value of the meta-stability bound on
$\lambda_4$.  For the space-time volume we can take the Hubble volume
and time,
\begin{equation}
V_4\sim \frac{1}{H^4_0},
\end{equation}
where $H_0$ is today's value of the Hubble constant. Longevity of the
vacuum requires that the tunneling probability in the volume $V_4\sim
1/H^4_{0}$ is smaller than $1$.  For $R$ between $R=1/(10^{10}~\gev)$
and $R=1/(10^{19}~\gev)$ this corresponds to a bounce action
$S_\text{E}\gtrsim 400-500$.  Therefore, as long as
\begin{equation}
|\lambda_4|\leq 0.052
\end{equation}
we are on the safe side. In the presence of higher-dimensional
operators, this inequality will be relaxed, and thus represents a very
conservative upper limit.\bigskip

A similar argument can be made when the higher dimensional operator is
actually destabilizing, \ie, $\lambda_6<0$.  In this a suitable
approximate potential is
\begin{equation}
V_\text{approx}(\varphi)=\frac{\lambda_4^\text{min}}{4} \, \varphi^4,
\qquad \text{with} \qquad 
\lambda^\text{min}_4=\lambda_4+ \frac{\lambda_6}{2k^2} \, \varphi_\text{max}^2 \; ,
\end{equation}
where $\varphi_\text{max}$ is the value where the potential is once
again stabilized by even higher dimensional terms.  Using the same
arguments as above the limit becomes
\begin{equation}
|\lambda^\text{min}_4|
=\left|\lambda_4+\frac{\lambda_6}{2k^2}\varphi_\text{max}^2\right|
\leq 0.052.
\end{equation}
We conclude, in accordance with Ref.~\cite{branchina} that the
presence of higher-order operators could have a significant impact on
the lifetime of the electroweak vacuum in a meta-stable situation, for
example if $\lambda_6$ is negative.  Estimates relying solely on
$\lambda_4$ strictly only apply to a restricted set of UV-completions. Large higher dimensional operators, in particular ones that destabilize the potential can have significant effects.

\section{Functional renormalization group}
\label{app:compare}

The $\beta$-functions for the running couplings used in this work
follow naturally from the functional renormalization group
(RG). Formulated in terms of a flow equation for a
momentum-scale-dependent effective action $\Gamma_k$, the functional
RG provides for efficient means to follow the flow of
higher-dimensional operators, to account for threshold effects and to
even approach regimes of strong coupling. Of course, the universal
one-loop $\beta$-functions dominantly used in the main text follow as
a simple by-product in the weak-coupling limit.

In the following, we briefly recall the essentials of the functional
RG, concentrating on its application to the present model. As this
method has already been employed in the present context to the pure
$\mathbb{Z}_2$ Higgs--Yukawa model~\cite{gies2013}, we focus on
the generalizations for the $SU(3)$ gauged version. Finally, we detail
how the universal one-loop $\beta$-functions arise by neglecting
higher-loop terms as well as RG improvement and by considering the
`deep Euclidean region'.\bigskip

The functional RG is a manifestation of the Wilsonian idea of solving
a quantum field theory by integrating out fluctuations in the path
integral, momentum shell by momentum shell. In its modern version, it
can be conveniently formulated in terms of a scale-dependent action
functional $\Gamma_k$. This is constructed such that it interpolates
between the UV-theory parameterized at the cutoff scale $\Lambda$ in
terms of the $\Gamma_{k\to\Lambda}\to S_\Lambda$ and the full
effective action in the IR, $\Gamma_{k\to0}\to
\Gamma_{\text{eff}}$. The latter contains, for instance, the full
effective potential as a local term without derivatives of the fields,
$\Gamma_{\text{eff}}=\int d^4 x \Big( V_{\text{eff}}+
\mathcal{O}((\partial \phi)^2) \Big)$. The change of the effective
action as a function of the RG scale $k$ (RG flow) is governed by the
Wetterich equation~\cite{christof_eq}
\begin{alignat}{5}
 \partial_t \Gamma_k = \frac{1}{2} \text{STr}\, \left[ \frac{\partial_t R_k}{\Gamma_k^{(2)}+R_k} \right], \qqquad t=\ln\frac{k}{\Lambda} \; ,
 \label{eq:WetterichEq}
\end{alignat}
where $\Gamma_k^{(2)}$ denotes the second derivative of $\Gamma_k$
with respect to the fluctuating fields ($H$, $\psi$, $\bar\psi$,
$A_\mu$, $\cdots$); the super-trace includes a minus sign for
fermionic trace parts.  It also involves a summation over the
eigenvalues of the regularized propagator, $(\Gamma^{(2)}+R_k)^{-1}$,
which in our case reduces to a momentum integration.  A new technical
ingredient is given by the regulator $R_k$ which can be thought of as
a momentum-dependent and $k$-dependent mass term. Specifying $R_k$
corresponds to specifying the details of the momentum shells
regularization. Different choices of $R_k$ correspond to different
regularization schemes.

Even though the Wetterich equation has a simple one-loop structure, it
is an exact equation as there exists an exact equivalence to the full
functional integral. The key ingredient for this is that the
propagator in the loop $\sim (\Gamma^{(2)}_k+R_k)$ denotes the full
propagator at the scale $k$. From the full solution of
Eq.\eqref{eq:WetterichEq}, for instance, all correlation functions
can be computed.\bigskip

While a perturbative expansion of the Wetterich equation reproduces
perturbation theory to any order, the flow equation can also be used
to extract non-perturbative information~\cite{rg_reviews}. Following
the Wilsonian idea, the scale dependent action $\Gamma_k$ contains not
just relevant or marginal couplings, but encodes effects of high-scale
quantum fluctuations in the presence of higher-order operators. For
practical calculations, this infinite series of operators in the
effective action is truncated. In the present context, a useful
ordering scheme defining our approximation of the exact flow is
given by an expansion of the effective action in terms of derivatives
of the field. To next-to-leading order this gives
\begin{equation}
\Gamma_k= \int d^4x \left[ V + \frac{Z_H}{2} (\partial_\mu H)^2 + \sum_{j=1}^{n_f} Z_{\psi_j} \bar\psi_j i\slashed{D} \psi_j + i \frac{1}{\sqrt{2}} \sum_{j=1}^{n_y}{\bar y}_j H \bar\psi_j\psi_j +\frac{Z_{G}}{4}F_{\mu \nu}^aF^{a\mu \nu} \right], \label{eq:trunc}
\end{equation}
where the potential $V$, the (bare) Yukawa couplings ${\bar y}_j$ as
well as all wave function renormalizations $Z_{H,\psi,G}$ are
$k$-dependent. The gauge-part of the action is also supplemented by a
gauge-fixing and Fadeev-Popov ghost contribution.

Inserting this ansatz into the Wetterich equation and expanding both
sides in terms of this basis of operators leads to the
$\beta$-functions, \ie the flow equations for the Yukawa couplings and
the wave function renormalizations. For the flow of the effective
potential, we obtain a $\beta$-functional defining a $\beta$-function
for every field value $H$. For clarity, let us write down this flow of
the effective potential explicitly:
\begin{alignat}{5}
\beta_{V}(H)= \frac{d\, V(H)}{d\, \log k} =&  \frac{k^4}{32 \pi^2} \left[ \frac{1- \frac{\eta_H}{6}}
{1 + \frac{V''(H)}{k^2 Z_H}}
- 4 \sum_{j=1}^{n_y} N_\text{c}  \frac{1-\frac{\eta_{\psi_j}}{5}}
{1+ \frac{{\bar y}_j^2H^2}{2k^2 Z_{\psi_j}^2}}  \right] \; .  
\label{eq:potflow} 
\end{alignat} 
Here, we have introduced the anomalous dimensions of the fluctuation
fields, defined by the wave function renormalization flow,
\begin{equation}
\eta_H=- \frac{d\, \log Z_{H}}{d\, \log k}, \quad \eta_{\psi_j}=-\frac{d\, \log
Z_{\psi_j}}{d\, \log k}, \quad \eta_G=- \frac{d\, \log Z_{G}}{d\, \log k}. \label{eq:etadef} 
\end{equation}
In Eq.\eqref{eq:potflow}, we have also made an explicit choice for the
regulator function in terms of the linear regulator
\cite{Litim:2001up}. Other choices lead to a qualitatively similar
behavior on the right hand side. This expression clearly contains
threshold effects, where for example the scalar contribution will be
dampened once the scalar mass, related to $V''$, becomes of order $k$.

From Eq.\eqref{eq:potflow}, the $\beta$-functions for the scalar
self-interactions $\lambda_{2n}$ can be obtained straightforwardly. To
do this, we expand the potential in powers of the field,
\begin{alignat}{6}
V_\text{eff}(k) =
     \frac{\bar\mu{(k)}^2}{2} H^2
   + \sum_{n=2} \frac{\bar\lambda_{2n}(k)}{k^{2n-4}} \; \left( \frac{H^2}{2} \right)^n \; .
\label{eq:potexp}
\end{alignat}
and introduce the renormalized couplings,
\begin{equation}
\mu^2=\frac{\bar\mu^2}{Z_H}, \quad \lambda_{2n}= \frac{\bar\lambda_{2n}}{Z_H^n}, \quad y=\frac{{\bar y}_j}{Z_{\psi_j} Z_H^{1/2}}, \quad g_s^2= \frac{\bar{g}_s^2}{Z_G}.
\end{equation}
Here, the rescaling with the wave function renormalizations
compensate for the field rescalings which are necessary to bring
Eq.\eqref{eq:trunc} to a canonical form, e.g., $H\to H/Z_H^{1/2}$,
etc. Inserting the expansion of Eq.\eqref{eq:potexp} into
Eq.\eqref{eq:potflow}, and expanding both sides to fourth order in
the field, we can read off the flow of $\lambda_4$,
\begin{equation}
\beta_{\lambda_4} = 2 \eta_H \lambda_4 + \frac{1}{16\pi^2} \left[ 9 \lambda_4^2 \frac{ 1 - \frac{\eta_H}{6} }{ (1 + \frac{\mu^2}{k^2})^3 } - \frac{15}{2} \lambda_6 \frac{ 1 - \frac{\eta_H}{6} }{ (1 + \frac{\mu^2}{k^2})^2 } \right] - \frac{1}{8\pi^2} \sum_{j=1}^{n_y} N_\text{c}y^4 \left( 1-\frac{\eta_{\psi_j}}{5} \right) .
\end{equation}
We observe that the scalar loop terms $\sim \lambda_4^2$ and $\sim
\lambda_6$ are suppressed if the mass parameter is larger than the RG
scale $k$. This signals a typical threshold effect characterizing the
decoupling of massive modes. In the present application, however,
$\mu^2$ is chosen such that it remains small for all values of $k$ and
eventually drops below zero near the weak scale $k_\text{EW}$. Then
all modes become massive and decouple, which can be made explicit in
the flow by expanding about $H=v$.\bigskip

For the high-energy applications discussed in the main text, we take
the limit of the deep Euclidean region, $\mu^2/k^2 \to 0$, as is
standard in perturbative computations. Using the one-loop result for
the anomalous dimension $\eta_H$ (also obtained from the flow equation
by studying the flow of the scalar kinetic term), $\eta_H= \sum_{j=1}^{n_y}
N_{\text{c}} y_j^2/(4\pi)^2$, we obtain the $\beta_{\lambda_4}$ function
as given in Eq.\eqref{eq:betas_toy}, except for the last term $\sim
g_F^4$ modelling the electroweak gauge sector which is described in
the main text. In fact, the ansatz of Eq.\eqref{eq:trunc} 
also yields higher-order terms like $\sim\lambda_4^2 \eta_H$,
corresponding to contributions from higher loops. The fact that these
terms arise from the one-loop formula in Eq.\eqref{eq:WetterichEq} is
due to the `RG-improvement' of the propagator in the loop. 

For the perturbative estimates discussed in the main text, these
higher-loop contributions are neglected, cf. Eq.\eqref{eq:betas_toy},
whereas they are fully included in the FRG results of
Fig.~\ref{fig:lambda6cutoff} further outlined below.

Similarly, $\beta_{\lambda_6}$ of Eq.\eqref{eq:betas_toy} is obtained
from an expansion of Eq.\eqref{eq:potflow} to order $H^6$, taking the
deep-Euclidean limit and ignoring RG-improvement terms. Also a
contribution $\sim \lambda_8$ is ignored in Eq.\eqref{eq:betas_toy},
whereas it is included in the FRG flow studies, see below.  The
derivation of $\beta_y$ of Eq.\eqref{eq:betas_toy} with $y_j\to y$ for
$n_y=1$ follows the same pattern. In the general case, the flow of the
Yukawa couplings for different fermions can be derived in this way;
then, the different fermions also acquire separate anomalous
dimensions. The standard one-loop $\beta_{g_s}$ function for the gauge
sector can most easily be derived from the flow equation, using the
background-gauge formalism. Here, the running of $g_{s}$ is directly
related to the anomalous dimension, $\beta_{g_s}=\eta_G g_s/2$ see
e.g.~\cite{rg_reviews}.

The manner by which we extracted the one-loop $\beta$-functions from
the Wetterich equation follows a general pattern: first expand both
sides in terms of the operators under consideration; this yields the
generally non-perturbative $\beta$-function of the corresponding
coupling within a given ansatz. For the one-loop order, all
RG-improvement terms corresponding to higher-loop corrections can
simply be dropped. The universal one-loop coefficients arise in the
limit of the deep Euclidean region which is the expected necessary
prerequisite for one-loop universality. While our explicit examples
discussed above have used the linear regulator as a special choice, it
can be proved that the one-loop coefficients for the renormalizable
operators are manifestly scheme independent. However, the
$\beta$-function coefficients for $\lambda_6$ do depend on the choice
of the regulator. This scheme dependence is physical in the sense that
it parameterizes how the Standard Model as an effective theory is
embedded into an underlying theory.\bigskip

For the non-perturbative functional RG results in the main text we
integrate the $\beta$-functions for the full ansatz
Eq.\eqref{eq:trunc}, including the mass parameter $\mu^2(k)$, all
threshold dependences and all RG-improvement terms due to the
back-feeding of anomalous dimensions. For the flow of the effective
potential we use a polynomial expansion to order $H^8$.  Higher orders
can easily be included, but do not lead to any significant change of
results. For the pure $\mathbb{Z}_2$ model, the local convergence has
been checked to order $\sim \lambda_{40}H^{40}$~\cite{gies2013}.

The flow equations for the pure $\mathbb{Z}_2$ model are given in
Refs.~\cite{gies2013,Gies:2009hq} for a general regulator function and
the linear regulator. To match them to the present matter content, all
fermion loop contributions have to be multiplied by a factor
$\sum_{j=1}^{n_y} N_c$, corresponding to the number of
fermions interacting with the scalar sector. In the following, we
confine ourselves to listing only the new contributions induced by the
$SU(N_c)$ gauge sector. The electroweak sector will be modelled in
terms of a fiducial coupling, as in Eq.\eqref{eq:betas_toy}.

Whereas the form of the flow of the effective potential in
Eq.\eqref{eq:potflow} or Eq.(12) in Ref.~\cite{gies2013} remains
unaffected, the flow of the Yukawa coupling receives direct
contributions from the gauge sector. Also for numerical purposes, it
is convenient to write this $\beta$-function as
\begin{equation}
\frac{d\, y_j^2}{d\, \log k} 
=  2\times \text{Eq.(13) of~\cite{gies2013}} 
+ \frac{1}{4\pi^2} \frac{N_c^2-1}{2N_c} g_s^2y_j^2 (3+\xi) \Big[ 2\kappa y_j^2 l_{2,1}^{(FB)4}(\kappa y_j^2,0;\eta_{\psi_j},\eta_{G}) 
- l_{1,1}^{(FB)4}(\kappa y_j^2,0;\eta_{\psi_j},\eta_G)\Big]
\, .
\label{eq:yq}
\end{equation}
The factor of 2 in front of the first term is due to different
conventions for the Yukawa coupling.  The dictionary for the
conventions used in this work and in Ref.~\cite{gies2013} reads:
$y_j^2\hat{=}2h^2$, $H^2/Z_H\hat{=}2k^2 \tilde{\rho}$, $V(H)\hat{=}k^4
u(\tilde\rho)$, where the LHS corresponds to this work and the RHS
to~\cite{gies2013}. The dimensionless quantity $\kappa$ is related to
the expansion point of the potential in field space; in the present
work, we have $\kappa=0$ at high-energy scales and $\kappa=v^2/(2k^2)$
in the broken regime near the weak scale. The threshold functions $l$
(and $m$ below) describe the (regulator-dependent) decoupling of
massive modes; for any admissible regulator, they approach finite
constants for vanishing first arguments and decrease to zero for large
first arguments. The result holds for arbitrary Lorenz-gauge parameter
$\xi$. All numerical studies have been performed in the Landau gauge
$\xi=0$.

Correspondingly, the fermion anomalous dimension reads
\begin{equation}
\eta_{\psi_j}= \text{Eq.(16) of~\cite{gies2013}} + 
\frac{1}{16\pi^2} \frac{N_c^2-1}{2N_c} g_s^2 \Big[ (3-\xi)m_{1,2}^{(FB)d}(\kappa y_j^2,0;\eta_G) - 3(1-\xi)\tilde{m}_{1,1}^{(FB)d}(\kappa y_j^2,0;\eta_G)  \Big]
, \label{eq:etapsi}
\end{equation}
while the corresponding equation for the scalar anomalous dimension
$\eta_H$ remains form invariant to Eq.(15) of Ref.~\cite{gies2013}.

In practice, the resulting system of coupled differential equations is
solved numerically. For stability reasons in the presence of
higher-dimensional operator, the flow is solved from the UV down to
the weak scale. The boundary conditions in the UV are chosen
implicitly such that the infrared observables are matched as described
below.

\section{Mapping between flow trajectories}
\label{app:mapping}

To study the RG evolution of the Standard Model, its physical masses
and couplings have to be related to the fundamental parameters
appearing in the $\beta$-functions.  Within the perturbative setting
this procedure is well-established up to NNLO~\cite{giudice}.  For our
perturbative results we follow these prescriptions at the
corresponding loop order, setting the parameters at the universal RG
scale $k=m_t^{\text{(pole)}}$ such that they reproduce a running top
mass $m_t=164~\gev$, a running Higgs mass $\mh=125~\gev$, and the
strong coupling $\alpha_S(m_Z)=0.1184$.  We list the explicit numbers
employed in our calculations for our model with $n_y=1$ in
Tab.~\ref{tab:parameters}.\bigskip

\begin{table}[!b]
\begin{tabular}{c | c c c}
\hline
 & $\lambda_4$ & $y$ & $g_s$\\ 
\hline
$k=m_t^{\text{(pole)}}$ & 0.129 & 0.940 & 1.167\\
$k=1$~TeV & 0.101 & 0.867 & 1.060\\
\hline
\end{tabular}
\caption{Values of the model parameters at the top pole mass and the
  RG matching scale $k=1$~TeV for the matching between the
  perturbative RG and the functional RG trajectories.}
\label{tab:parameters} 
\end{table}

\begin{figure}[t]
\includegraphics[width=1.0\textwidth]{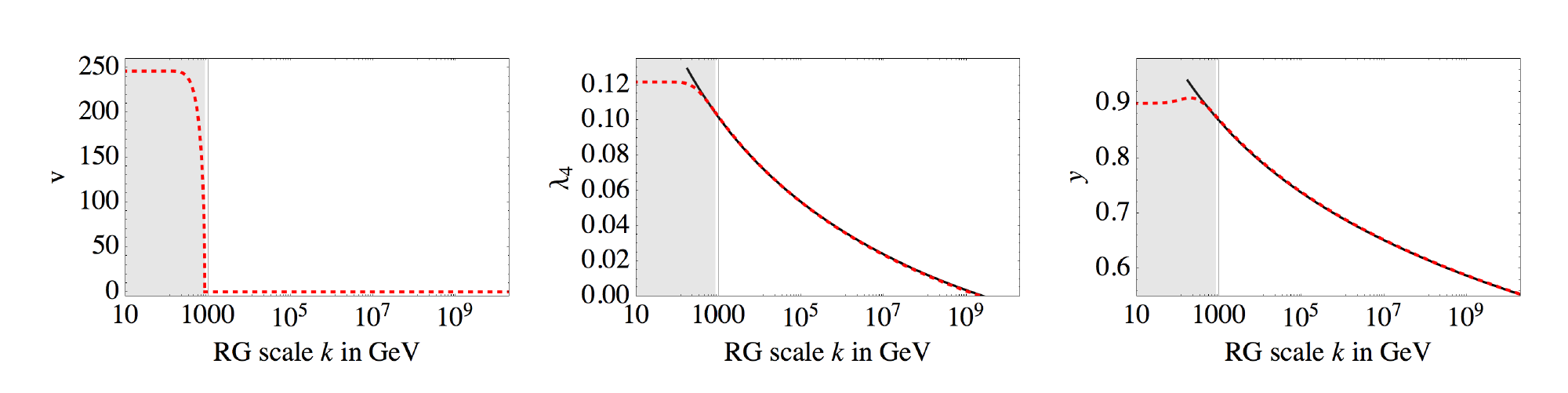}
\caption{Comparison and matching of FRG flows with the perturbative
  setting. The left panel distinguishes the symmetric regime and the
  regime of spontaneously broken symmetry. The center and right panels
  show the FRG flows (red dotted) as well as the perturbative flows
  (black curve) of the Higgs coupling and the top Yukawa coupling. The
  running couplings are matched at $1~\tev$.}
\label{fig:compareFRGandPT} 
\end{figure}

The functional RG approach allows us to access the regime of
spontaneously broken symmetry, where a non-vanishing minimum
$H_{\text{min}, k}\neq 0$ of the effective potential appears in the RG
flow. This way all particle masses are generated dynamically. In
contrast to the perturbative setting, the running minimum enters the
RG evolution of the Higgs quartic coupling and the Yukawa coupling
explicitly: The generated running masses $m_t(k) \propto y
H_{\text{min}, k}$ and $\mh(k) \propto \sqrt{\lambda_4} H_{\text{min},
  k}$ induce a threshold behavior that leads to a freeze-out of model
runnings in the infrared, as shown in Fig.~\ref{fig:compareFRGandPT}.
The FRG flow for the RG relevant direction parameterized by the mass
parameter $\mu^2$ has to be fine-tuned at the UV-cutoff scale, such
that the minimum of the effective potential approaches the SM value of
the vev in the infrared $H_{\text{min}, k\rightarrow 0}=v$. This
fine-tuning is a manifestation of the hierarchy problem, which is
dealt with here by an explicit choice of UV-parameters.

The threshold behavior in the FRG is non-universal, which means it
depends on the choice of regulator function. This dependence is
natural as it parametrizes the details of decoupling of massive modes
from the flow. In the symmetry-broken regime, the threshold behavior
leads to a significant deviation of the running couplings from their
values in the perturbative setting, which is strictly tied to the
symmetric deep Euclidean region; the latter is, of course,
inappropriate for the threshold region.  To compare to the
perturbative results and extract universal physical results, we have
to establish a matching procedure.  We choose to match the flows at
the scale 1~TeV, which is sufficiently deep inside the Euclidean
region but not yet affected by possible higher-dimensional operators,
cf.~Tab~\ref{tab:parameters}.  

Our procedure equates the running couplings in the perturbative
$\msbar$ scheme and the FRG scheme at the matching scale.  Approaching
that scale from the ultraviolet, typical FRG trajectories are still in
the symmetric regime, an electroweak minimum of the potential has not
been generated yet, and the threshold effects in the FRG
$\beta$-functions are subleading. In Fig.~\ref{fig:compareFRGandPT} we
see that this regime features flows of the Higgs self-coupling and the
top Yukawa coupling closely resembling the perturbative RG flows.  As
the values for the couplings agree with those of the $\msbar$ scheme
at the matching scale, our matching scheme guarantees that UV-initial
conditions for the FRG flow are mapped to physical values in the
infrared with high accuracy. Qualitatively, this mapping works already
at the level of our FRG trajectories without a matching procedure,
since threshold effects automatically provide for a freeze-out of the
RG flow, such that in principle physical values can be extracted at
$k=0$.  In a more elaborate FRG setup, using a more sophisticated
truncation of the operator space, our mapping using $\msbar$
parameters will become obsolete and quantitatively precise values for
physical observables can be read off from the FRG flow trajectories at
$k=0$ directly.

Our choice of matching scale is also applicable when the
symmetry-breaking scale lies at around $1~\tev$, as threshold effects
remain small close to that scale.  Matching the flows at $1~\tev$
therefore provides us with a well-defined procedure to compare the two
RG schemes.  As an added benefit we can thoroughly study the FRG
evolution of our toy model consistently including higher-dimensional
operators and threshold effects in a Standard Model context.

\section{Modelling electroweak contributions to the full effective potential}
\label{advancedfudge}

In the present work, the electroweak contributions are modelled by
means of a fiducial gauge coupling contributing to the perturbative
$\beta$-functions Eq.~\eqref{eq:betas_toy}. In particular, we neglected 
a similar contribution to the flow of $\lambda_6$ in this simple model. 
With the full $\beta$-functional for the full effective potential
Eq.~\eqref{eq:potflow} at hand, we can in fact model the electroweak
gauge contributions to the full potential including the flow of the
mass-parameter $\mu^2(k)$ as well as all higher couplings
$\lambda_{6,8,\dots}(k)$. To lowest order, we add the contribution of
a fiducial gauge loop to the $\beta$-functional \cite{gies},
\begin{equation}
\text{Eq.~\eqref{eq:potflow}} \quad \to \quad \text{Eq.~\eqref{eq:potflow}} 
+ \frac{k^4}{32 \pi^2} \frac{c_V}{1 + \frac{g_F^2 H^2}{4 k^2}},
\label{eq:potflow:adv}
\end{equation}
with the fiducial gauge coupling $g_F$ and a numerical constant $c_V$. 
As in the main text, we choose these fiducial parameters in such a way 
that the running of the quartic coupling $\lambda_4$ is quantitatively 
similar to the running of the Standard Model Higgs quartic coupling, cf. 
Eq.~\eqref{eq:betas_toy} and Fig.~\ref{fig:modelflows}. Comparing to 
Eq.~\eqref{eq:betas_toy}, we identify $c_V = 16\, c_{\lambda} = 9$. This 
choice leads to the same flow equation for the quartic coupling as in the 
main text. In addition, also the flow equations of all other couplings in 
the scalar sector ($\mu^2(k)$, $\lambda_{6,8,...}(k)$) receive contributions 
proportional to an appropriate power of the fiducial gauge coupling $g_F$. 
The flow of all scalar couplings 
is only slightly modified such that the 
solution 
for the simple model lies almost on top of the results with the 
advanced fiducial modification Eq.~\eqref{eq:potflow:adv}. The only relevant 
quantitative change is the flow of the mass parameter $\mu^2(k)$. 
The additional bosonic degrees of freedom reduce the ratio $\mu^2(k)/k^2$ on a wide range of scales.

The main modification introduced by the advanced model Eq.~\eqref{eq:potflow:adv} 
affects the validity regime of the unique-minimum condition Eq.~\eqref{eq:stability}. 
This leads to a shift of the border between region II and III of 
Fig.~\ref{fig:lambda6cutoff}. Smaller values for $\mu^2(\Lambda)$ have to be 
compensated by a smaller UV cutoff $\Lambda$ for fixed $\lambda_4(\Lambda)$ and 
$\lambda_6(\Lambda)$, resulting in a depletion of the transition curve. 
This effect is less pronounced at the transition between region I and II because the relative reduction in $\mu^2/k^2$ is smaller.

\begin{figure}[t]
\includegraphics[width=0.5\linewidth]{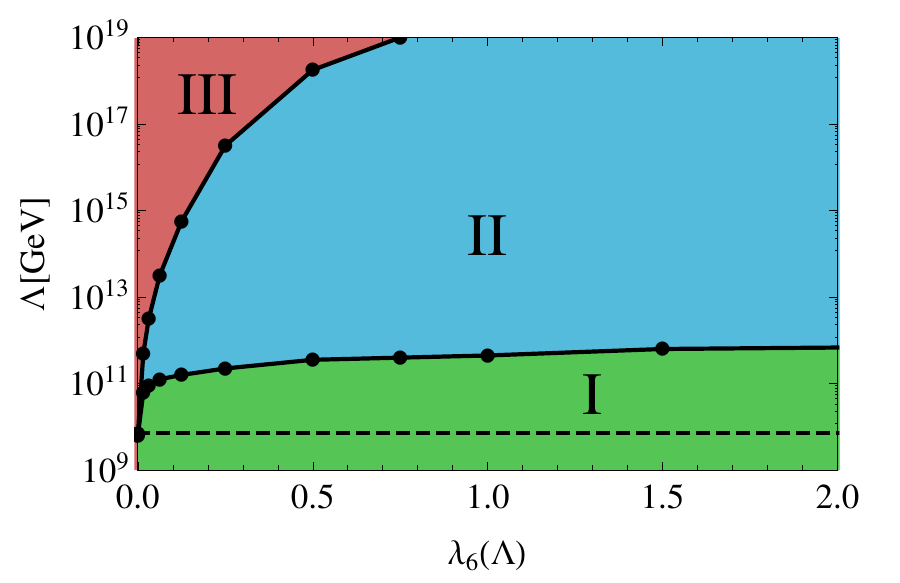}
\caption{Different stability regions as a function of $\Lambda$ and
  $\lambda_{6}$ analogous to Fig.~\ref{fig:lambda6cutoff} for the
  simple model. In region~I
  (green)
  the potential is stable everywhere;
  in region~II
  (blue)
  the UV-potential is stable, while the potential
  is only pseudo-stable for intermediate scales $v < k < \Lambda$; in
  region~III  
  (red)
  the UV-potential violates the unique-minimum
  condition. The inclusion of electroweak contributions to the full potential leads only to minor modifications of the border between region I and II. Region III becomes somewhat larger as the smaller mass parameter $\mu^2$ facilitaes a violation of the unique-minimum condition already for smaller cutoffs.}
\label{fig:lambda6cutoff:adv}
\end{figure}


\end{document}